\documentclass[10pt,final,journal]{IEEEtran}
\usepackage{cite}

\ifCLASSINFOpdf
\else
\fi
\usepackage{amsmath}
\usepackage{algorithmic}
\usepackage{bm}
\usepackage{subfigure}
\usepackage{wasysym}
\usepackage{amssymb}
\usepackage{array}
\usepackage{graphicx}
\graphicspath{{../Figures/Biography Photos/}{../jpeg/}}
\usepackage{soul}
\usepackage{gensymb}

\usepackage[flushleft]{threeparttable}
\usepackage{hhline}
\usepackage[space]{grffile}
\usepackage{marvosym}

\usepackage{comment}
\usepackage{float}

\usepackage{lipsum}
\usepackage{booktabs}
\usepackage{adjustbox}

\usepackage{tabularx}
\usepackage{siunitx}
\usepackage{ragged2e}

\begin{document}
%
\title{Adaptive and Dynamically Constrained Process Noise Estimation for Orbit Determination}
%
%
%

\author{Nathan~Stacey,
        Simone~D'Amico,
\thanks{N. Stacey and S. D'Amico are with the Department of Aeronautics and Astronautics, Stanford University, Stanford, CA, 94305 USA (email: nstacey@stanford.edu, damicos@stanford.edu)}
\thanks{This article has been accepted for publication in \textit{IEEE Transactions on Aerospace and Electronic Systems}, DOI: 10.1109/TAES.2021.3074205. Copyright © 2021 IEEE. Personal use of this material is permitted. Permission from IEEE must be obtained for all other uses, in any current or future media, including reprinting/republishing this material for advertising or promotional purposes, creating new collective works, for resale or redistribution to servers or lists, or reuse of any copyrighted component of this work in other works.}}

\maketitle

\begin{abstract}
This paper introduces two new algorithms to accurately estimate the process noise covariance of a discrete-time Kalman filter online for robust orbit determination in the presence of dynamics model uncertainties. Common orbit determination process noise techniques, such as state noise compensation and dynamic model compensation, require offline tuning and a priori knowledge of the dynamical environment. Alternatively, the process noise covariance can be estimated through adaptive filtering. However, many adaptive filtering techniques are not applicable to onboard orbit determination due to computational cost or the assumption of a linear time-invariant system. Furthermore, existing adaptive filtering techniques do not constrain the process noise covariance according to the underlying continuous-time dynamical model, and there has been limited work on adaptive filtering with colored process noise. To overcome these limitations, a novel approach is developed which optimally fuses state noise compensation and dynamic model compensation with covariance matching adaptive filtering. This yields two adaptive and dynamically constrained process noise covariance estimation techniques. Unlike many adaptive filtering approaches, the new techniques accurately extrapolate over measurement outages and do not rely on ad hoc methods to ensure the process noise covariance is positive semi-definite. The benefits of the proposed algorithms are demonstrated through two case studies: an illustrative linear system and the autonomous navigation of two spacecraft orbiting an asteroid.
\end{abstract}


%
\IEEEpeerreviewmaketitle

\section{Introduction}
%
%
%
%
\IEEEPARstart{I}{n} orbit determination, there are always differences between the modeled and true spacecraft dynamics due to complex forces that cannot be modeled perfectly. These numerous forces include gravity, solar radiation pressure, atmospheric drag, third-body perturbations, tidal effects, and propulsive maneuvers. Furthermore, reduced order dynamics models are often used for onboard navigation due to computational limits. In Kalman filtering, the uncertainty of the filter dynamics model is captured through the modeled process noise. Inaccurate process noise models can lead to large estimation error as well as filter inconsistency and divergence\cite{schlee_divergence_1967,mehra_approaches_1972}. Modeling process noise for asteroid missions is especially challenging because the dynamical environment is poorly known a priori, and the process noise statistics can change significantly as the spacecraft transitions between high and low altitude orbits. It is increasingly difficult when there is limited human intervention such as in the Autonomous Nanosatellite Swarming (ANS)\cite{ANS,stacey_autonomous_2018,lippe_spacecraft_2019} mission concept, which utilizes an autonomous swarm of small spacecraft to characterize an asteroid.

Two common approaches to modeling process noise for orbit determination in a discrete-time Kalman filtering framework are state noise compensation (SNC) and dynamic model compensation (DMC), which have been succesfully employed in a variety of scenarios \cite{schutz_statistical_2004,sullivan_nonlinear_2017,damico_autonomous_2010,giralo_distributed_2019,montenbruck_reduced_2005,bierman_factorization_1977,tapley_orbit_1973,myers_dynamical_1975,leonard_gravity_2013}. These methods explicitly attribute spacecraft state process noise to unmodeled accelerations, which are the difference between the true and modeled accelerations. SNC treats the unmodeled accelerations as continuous-time process noise, which is modeled as a zero-mean white Gaussian process with known power spectral density. However, in reality unmodeled accelerations are correlated in time. DMC allows for time correlation of the unmodeled accelerations by augmenting the state with empirical accelerations. The dynamics of the empirical accelerations are frequently modeled as a first-order Gauss-Markov process that treats the unmodeled accelerations as exponentially correlated in time. DMC can provide higher orbit determination accuracy than SNC\cite{myers_dynamical_1975,tapley_orbit_1973,goldstein_real-time_2001}. On the other hand, SNC is simpler to implement and less computationally expensive than DMC.

A major drawback of SNC and DMC is that time-intensive offline tuning is required to determine the process noise power spectral density\cite{cruickshank_genetic_1998,schutz_statistical_2004,goldstein_real-time_2001,leonard_gravity_2013}. The tuned power spectral density is no longer optimal when the process noise statistics change, which naturally occurs due to variations in space weather and spacecraft properties. The process noise statistics can also change whenever the orbit is altered as well as between periapsis and apoapsis. Moreover, SNC and DMC are poorly suited to scenarios where the dynamical environment is not well known a priori because the offline tuning cannot be accurately completed before the mission. Genetic model compensation was developed to adaptively tune the main diagonal of the process noise power spectral density for DMC through a genetic optimization algorithm\cite{cruickshank_genetic_1998,goldstein_real-time_2000}. However, genetic model compensation is not widely used due to its complex implementation and sensitivity to multiple hyperparameters that require offline tuning.

Alternatively, the process noise covariance may be estimated through adaptive filtering techniques, which are commonly divided into four categories: Bayesian\cite{alspach_parallel_1974,lainiotis_optimal_1970,wiberg_online_2000,sarkka_recursive_2009,friedland_estimating_1982,li_recursive_1994}, maximum likelihood\cite{kashyap_maximum_1970,shumway_approach_1982}, correlation\cite{odelson_new_2006,belanger_estimation_1974}, and covariance matching (CM)\cite{karlgaard_robust_2010,moghe_adaptive_2019,myers_adaptive_1976,gutman_tracking_1990}. An early survey on adaptive filtering techniques is given by Mehra\cite{mehra_approaches_1972}. A more recent and comprehensive survey is provided by Dun\' ik et al.\cite{dunik_noise_2017} with a focus on correlation methods. 

However, there are limitations to directly applying current adaptive filtering techniques to orbit determination. For example, approaches derived for linear time-invariant systems\cite{zhang_identification_2020,mehra_identification_1970,moghe_adaptive_2019,odelson_new_2006,mohamed_adaptive_1999}
have limited applicability. Additionally, ad hoc methods are often used to ensure the process noise covariance is positive semi-definite\cite{mohamed_adaptive_1999,fraser_adaptive_2021,
 myers_adaptive_1976,myers_filtering_1974,moghe_adaptive_2019}, which can lead to biased estimates. Many adaptive filtering techniques have a limited ability to extrapolate over measurement outages where the process noise covariance may be significantly different from previous values. Existing approaches also estimate the process noise covariance directly without constraining the estimate according to the underlying continuous-time dynamical model. Consequently, the estimated process noise covariance may not be realizable given that the process noise is ascribed to continuous-time stochastic unmodeled accelerations. Many correlation and maximum likelihood methods do not update the noise covariances used in the filter online or only do so infrequently. These techniques are referred to as feedback-free by Dun\' ik et al.\cite{dunik_noise_2017} and are ill-suited to online applications. Furthermore, in orbit determination the process noise is often highly correlated in time. However, there has been limited work on adaptive filtering for systems with colored process noise\cite{lee_state_2017}, which Dun\' ik et al.\cite{dunik_noise_2017} recently identified as a necessary area of future research. Although CM techniques suffer from many of these limitations, they are widely used because they are simple to implement and computationally efficient. Since many other adaptive filtering algorithms are not computationally tractable for typical spacecraft onboard processors, several authors have explored using CM for orbit determination\cite{myers_adaptive_1976,sullivan_nonlinear_2017,karlgaard_robust_2010,fraser_adaptive_2021}.

This paper overcomes the aforementioned limitations of SNC, DMC, and existing adaptive filtering techniques by optimally fusing SNC and DMC with CM through a constrained weighted least squares optimization. This yields two novel techniques for estimating the process noise covariance called adaptive SNC (ASNC) and adaptive DMC (ADMC). The adaptability of the proposed algorithms is a significant advantage over SNC and DMC. In contrast to many adaptive filtering approaches, ASNC and ADMC are well suited to onboard orbit determination because they are computationally efficient and do not assume a linear time-invariant system. The proposed techniques update the filter process noise covariance at each filter call, which is an important advantage in online applications over feedback-free methods. Additionally, the developed techniques are constrained according to the underlying continuous-time dynamical model and do not rely on ad hoc methods to ensure the process noise covariance is positive semi-definite. The new techniques also accurately extrapolate over measurement outages and can leverage a priori bounds on the process noise power spectral density. In particular, ADMC can provide more accurate estimation than ASNC for systems with colored process noise. In addition to Earth-based missions, ASNC and ADMC are applicable to asteroid missions such as ANS\cite{ANS,stacey_autonomous_2018,lippe_spacecraft_2019} where the dynamical environment is poorly known a priori, and the process noise statistics are time-varying. Another potential use for ASNC and ADMC is to tune the process noise power spectral density for missions that will use either SNC or DMC. 

The following section provides background on SNC, DMC, and CM. The next section describes how SNC and DMC are each optimally fused with CM through a constrained weighted least squares minimization to yield ASNC and ADMC. This includes a derivation of the weighting matrix used in the least squares minimization as well as discussion on efficiently solving the optimization problem. In the subsequent section, the advantages of ASNC and ADMC are demonstrated through an illustrative linear system. The benefits of the developed techniques are further substantiated in the penultimate section through a nonlinear system comprising the idealized navigation of two spacecraft orbiting an asteroid. Finally, conclusions are presented based on the results of the two case studies. 
 
\section{Background}
\subsection{State Noise Compensation}
Consider a linear time-varying system subject to process noise described by
\begin{equation}\label{eq:EOM}
	\dot{\bm{x}}(t) = \mathbf{A}(t)\bm{x}(t) + \mathbf{B}(t)\bm{u}(t) + \mathbf{\Gamma}(t)\bm{\epsilon}(t)
\end{equation}
Here $\mathbf{A}$ is the plant matrix, $\mathbf{B}$ is the control input matrix, $\bm{u}$ is the control input, $\mathbf{\Gamma}$ is the process noise mapping matrix, and $\bm{\epsilon}$ is the continuous-time process noise. The system state is denoted by $\bm{x}$ and contains either a Cartesian or orbital element spacecraft state in orbit determination. For a linear system, the only sources of spacecraft state process noise are unmodeled accelerations and numerical error. In SNC, the process noise is attributed entirely to unmodeled accelerations since they generally create orders of magnitude more process noise than numerical error. The unmodeled accelerations are described by $\bm{\epsilon}$, which is modeled as a zero-mean white Gaussian process\cite{myers_filtering_1974,schutz_statistical_2004}. The realization of each $\bm{\epsilon}$ is an acceleration, and the autocovariance of $\bm{\epsilon}$ is
\begin{equation}\label{eq:SNC Covariance}
	\text{E}[\bm{\epsilon}(t)\bm{\epsilon}(\tau)^T] 	= \mathbf{\widetilde{Q}}(t)\delta(t-\tau)		
\end{equation}
where $\mathbf{\widetilde{Q}}(t)$ is the process noise power spectral density, and $\delta(\cdot)$ is the Dirac delta function. The solution to Eq. (\ref{eq:EOM}) can be written as
\begin{equation}\label{eq:true dynamics} 
	\bm{x}_k = \mathbf{\Phi}_{k}\bm{x}_{k-1} + \int_{t_{k-1}}^{t_k} \mathbf{\Phi}(t_k,\tau)\mathbf{B}(\tau)\bm{u}(\tau)d\tau +  \bm{w}_k
\end{equation}
where $\bm{x}_k$ is the state at time step $k$, $\mathbf{\Phi}_{k} = \mathbf{\Phi}(t_k,t_{k-1})$ is the state transition matrix that propagates the state from time $t_{k-1}$ to $t_k$, and
\begin{equation}\label{eq:w int}
\bm{w}_k = \int_{t_{k-1}}^{t_k} \mathbf{\Phi}(t_k,\tau)\mathbf{\Gamma}(\tau)\bm{\epsilon}(\tau)d\tau
\end{equation}
is the discrete-time process noise. It is easily shown that $\bm{w}$ is uncorrelated in time and that $\bm{w}_k  \sim \mathcal{N}(0,\mathbf{Q}_{k})$ where the process noise covariance is\cite{myers_adaptive_1976,schutz_statistical_2004}
\begin{equation}\label{eq:Q}
	\mathbf{Q}_{k} = \int_{t_{k-1}}^{t_k}  \mathbf{\Phi}(t_k,\tau)  \mathbf{\Gamma}(\tau)  \mathbf{\widetilde{Q}}(\tau)  \mathbf{\Gamma}(\tau)^T  \mathbf{\Phi}(t_k,\tau)^T  d\tau
\end{equation}
In SNC, $\mathbf{\widetilde{Q}}$ is assumed constant and tuned offline. To facilitate tuning, the elements of $\bm \epsilon$ are generally assumed to be independent such that $\mathbf{\widetilde{Q}}$ is diagonal\cite{myers_adaptive_1976,goldstein_real-time_2001,schutz_statistical_2004}. The matrix $\mathbf{\widetilde{Q}}$ is assumed diagonal throughout this paper. The process noise covariance is used in the Kalman filter time update of the formal covariance, or error covariance, which is given by
\begin{equation}\label{eq:covariance time update}
		\mathbf{P}_{k|k-1} = \mathbf{\Phi}_{k}\mathbf{P}_{k-1|k-1}\mathbf{\Phi}_{k}^T+\mathbf{Q}_{k}
\end{equation}
Here $\mathbf{P}_{i|j}$ is the formal covariance at time step $i$ taking into account all the measurements through time step $j$. In general, the spacecraft dynamics are nonlinear. For a nonlinear system that has been linearized, the process noise covariance should be large enough to also accommodate errors due to dynamical nonlinearities in the propagation of the mean state estimate and associated error covariance. 

\subsection{Dynamic Model Compensation}
In reality, the unmodeled accelerations are correlated in time. DMC takes this into account by augmenting the state vector with empirical accelerations, which are also referred to in literature as ficticious or compensative accelerations\cite{cruickshank_genetic_1998}. Although higher order models may be used\cite{leonard_gravity_2013}, empirical accelerations are often modeled as the first-order Gauss-Markov process\cite{schutz_statistical_2004,tapley_orbit_1973}\ 
\begin{equation}\label{eq:ae dynamics}
	\dot{\bm{\tilde{a}}}(t) = -\bm{\beta}\bm{\tilde{a}}(t)+\bm{\epsilon}(t)
\end{equation}
which is a linear stochastic differential equation known as a Langevin equation\cite{cruickshank_genetic_1998}. Here $\bm{\tilde{a}}$ is a vector containing an empirical acceleration for each spatial dimension. Again, $\bm{\epsilon}$ is a zero-mean white Gaussian process whose autocovariance is given in Eq. (\ref{eq:SNC Covariance}). The matrix $\mathbf{\widetilde{Q}}$ is assumed constant, and both $\mathbf{\widetilde{Q}}$ and $\bm{\beta}$ are determined through offline tuning. To facilitate tuning, it is typically assumed that  the empirical accelerations are independent from one another such that $\mathbf{\widetilde{Q}}$ and $\bm{\beta}$ are diagonal\cite{cruickshank_genetic_1998,tapley_orbit_1973,goldstein_real-time_2001,myers_dynamical_1975}. The offline tuning can be performed manually through trial and error\cite{goldstein_real-time_2001} or by fitting the empirical acceleration autocovariance model to sample accelerations\cite{leonard_gravity_2013}. Alternatively, $\bm{\beta}$ can be estimated as part of the state. However, when estimating $\bm{\beta}$ the performance of DMC is sensitive to the modeled process noise covariance of $\bm{\beta}$, which is tuned offline\cite{myers_dynamical_1975}. After determining $\mathbf{\widetilde{Q}}$ and $\bm{\beta}$, $\mathbf{Q}_{}$ is computed through Eq. (\ref{eq:Q}) and used in the time update of the formal covariance matrix in Eq. (\ref{eq:covariance time update}). Note that $\mathbf{\Phi}$ and $\mathbf{\Gamma}$ are different for DMC and SNC since DMC augments the state with empirical accelerations.

Assuming the empirical accelerations are independent and that $\mathbf{\widetilde{Q}}$ and $\bm{\beta}$ are constant, the solution to Eq. (\ref{eq:ae dynamics}) is\cite{schutz_statistical_2004,bierman_factorization_1977}
\begin{equation}\label{eq:emp acc solution}
	\tilde{a}_{i}(t) = e^{-\beta_i(t-t_0)}\tilde{a}_{i}(t_0) + \int_{t_0}^{t} e^{-\beta_i(t-\tau)}\epsilon_i(\tau)d\tau
\end{equation}
where $\tilde{a}_{i}$ is the $i$\textsuperscript{th} component of $\bm{\tilde{a}}$, $\epsilon_{i}$ is the $i$\textsuperscript{th} component of $\bm{\epsilon}$, and $\beta_i$ is the $i$\textsuperscript{th} element of the main diagonal of $\bm{\beta}$. The first term in Eq. (\ref{eq:emp acc solution}) is deterministic and is included in the filter dynamics model. The second term is stochastic with a mean of zero and is taken into account through the process noise covariance. For a finite $\mathbf{\widetilde{Q}}$ and $\beta_i=0$, the model in Eq. (\ref{eq:emp acc solution}) reduces to a random walk process. On the other hand, the model approaches a zero-mean Gaussian white noise sequence as $\beta_i \to \infty$. Modeling the time correlation of the unmodeled accelerations allows DMC to achieve greater estimation accuracy than SNC when optimal values of $\mathbf{\widetilde{Q}}$ and $\bm{\beta}$ are used\cite{myers_dynamical_1975,tapley_orbit_1973,goldstein_real-time_2001}. The direct estimate of the unmodeled accelerations may also be useful for improving the dynamical model in post-flight analyses\cite{myers_dynamical_1975,schutz_statistical_2004}. However, like SNC the required offline tuning makes it difficult to use DMC when the dynamical environment is poorly known a priori or the process noise statistics are time-varying.

\subsection{Covariance Matching Adaptive Filtering}
Consider the linear time-varying dynamical model shown in Eq. (\ref{eq:true dynamics}) with discrete-time measurements
\begin{equation}\label{eq:linear measurement model}
	\bm{z}_k = \mathbf{H}_{k} \bm{x}_k + \bm{\nu}_k
\end{equation}
Here $\mathbf{H}_{k}$ is the measurement matrix at time step $k$, and $\bm{\nu}_k$ is zero-mean Gaussian noise with autocovariance $\text{E}[\bm{\nu}_i\bm{\nu}_j^T]=\mathbf{R}_{i}\delta_{ij}$ where $\delta_{ij}$ is the Kronecker delta function. Many CM techniques, also referred to as innovation-based estimation, have been proposed to adaptively tune the process noise and measurement noise covariances by setting the theoretical innovation covariance equal to an empirical value\cite{mehra_approaches_1972}. One common method estimates $\mathbf{Q}_{k}$ through an average over a sliding window of length $N$ as
\begin{equation}\label{eq:Qhat estimate} 
	\mathbf{\hat{Q}}_{k} = \frac{1}{N}\sum_{p=k-N}^{k-1} \left(\mathbf{P}_{p|p}-\mathbf{\Phi}_{p}  \mathbf{P}_{p-1|p-1} \mathbf{\Phi}_{p}^T+\bm{\Delta}_{p}^x \bm{\Delta}_p^{x^T}\right)
\end{equation}
assuming $\mathbf{Q}$ is constant and an optimal filter. The state correction $\bm{\Delta}_{k}^x$ and the pre-fit residual or innovation $\bm{\Delta}_{k}^z$ are defined as 
\begin{align}
   \bm{\Delta}_{k}^x		&=\mathbf{K}_{k}\bm{\Delta}_{k}^z \label{eq:state innovation}\\
   \bm{\Delta}_{k}^z 	&= \bm{z}_k-\mathbf{H}_{k}\bm{\hat{x}}_{k|k-1}
\end{align} 
Here
\begin{equation}
 \mathbf{K}_{k} = \mathbf{P}_{k|k-1}  \mathbf{H}_{k}^T \mathbf{S}_{k}^{-1}
\end{equation}
is the Kalman gain, 
\begin{align}
\mathbf{S}_{k} &= \text{E}[\bm{\Delta}_{k}^z \bm{\Delta}_{k}^{z^T}] \\
							  &= \mathbf{H}_{k} \mathbf{P}_{k|k-1} \mathbf{H}_{k}^T + \mathbf{R}_{k}
\end{align}
is the innovation covariance, and $\hat{\bm{x}}_{i|j}$ is the mean state estimate at time step $i$ taking into account all the measurements through time step $j$. Eq. (\ref{eq:Qhat estimate}) was originally derived by Myers and Tapley\cite{myers_adaptive_1976,myers_filtering_1974}, with the difference being that here it is assumed that $\bm{w}_k$ is known to be zero-mean. Interestingly, Fraser\cite{fraser_adaptive_2021} shows that maximum likelihood estimation can also be used to obtain Eq. (\ref{eq:Qhat estimate}) whereas Mohamed and Schwarz\cite{mohamed_adaptive_1999} use maximum likelihood estimation to derive a similar result for linear time-invariant systems. 

An obvious shortcoming of Eq. (\ref{eq:Qhat estimate}) is that $\mathbf{\hat{Q}}_{k}$ may not be positive semi-definite. In order to be directly used in a Kalman filter, each $\mathbf{\hat{Q}}_{k}$ must be guaranteed positive semi-definite to be a valid covariance matrix and avoid losing the positive definiteness of the formal covariance in Eq. (\ref{eq:covariance time update}). In practice, ad hoc methods are used to ensure $\mathbf{\hat{Q}}_{k}$ is positive semi-definite. In the original derivation of Eq. (\ref{eq:Qhat estimate}), Myers and Tapley\cite{myers_adaptive_1976,myers_filtering_1974} suggest setting the diagonal of $\mathbf{\hat{Q}}_{k}$ equal to its absolute value. However, this results in a biased estimate and does not guarantee $\mathbf{\hat{Q}}_{k}$ is positive semi-definite. The most common approach is to assume the first two terms in Eq. (\ref{eq:Qhat estimate}) are negligible at steady state. Under this assumption, Eq. (\ref{eq:Qhat estimate}) reduces to\cite{fraser_adaptive_2021,mohamed_adaptive_1999}
\begin{equation}\label{eq:Qhat estimate ss}
	\mathbf{\hat{Q}}_{k} = \frac{1}{N}\sum_{p=k-N}^{k-1} \bm{\Delta}_{p}^x \bm{\Delta}_{p}^{x^T}
\end{equation}
which ensures $\mathbf{\hat{Q}}_{k}$ is positive semi-definite. However, steady state conditions only guarantee $\mathbf{P}_{p|p} = \mathbf{P}_{p-1|p-1}$, which is not a sufficient condition for $\mathbf{P}_{p|p}-\mathbf{\Phi}_{p}  \mathbf{P}_{p-1|p-1} \mathbf{\Phi}_{p}^T=\mathbf{0}$. More rigorously, a sufficient condition for the first two terms in Eq. (\ref{eq:Qhat estimate}) to add to zero is for the filter to be at steady state and the state transition matrix to be identity. Generally, the state transition matrix of an orbital element state remains close to identity for longer time steps than a Cartesian state because most orbital elements vary more slowly in time than position and velocity. However, if the estimated state also contains force model parameters or the spacecraft is in a highly perturbed orbit, the state transition matrix may only be close to identity for very small time steps, even for orbital element states. Moreover, the measurement rate may be limited by measurement availability or computational resources. Note that CM has a single tunable parameter, which is the length of the sliding window. This parameter should be chosen based on how quickly $\mathbf{Q}$ is expected to vary in time. A longer sliding window provides a more accurate estimate of $\mathbf{Q}$ when it is approximately constant. On the other hand, a shorter sliding window provides quicker adaptation of a time-varying $\mathbf{Q}$\cite{mohamed_adaptive_1999}. This CM technique assumes a steady state optimal filter, and there are no formal mathematical guarantees of convergence to the true $\mathbf{Q}$. However, Eq. (\ref{eq:Qhat estimate ss}) is computationally efficient, can update the value of $\mathbf{Q}$ used in the filter online, and is easily applied through an extended Kalman filter (EKF) or unscented Kalman filter (UKF) to nonlinear systems, which is typically the case in orbit determination. Note that several authors have successfully applied this CM approach to orbit determination\cite{myers_adaptive_1976,sullivan_nonlinear_2017,karlgaard_robust_2010,fraser_adaptive_2021}.
 
There are several drawbacks to directly applying the aforementioned CM technique to orbit determination. First, there are many systems for which the state transition matrix is not close to identity and Eq. (\ref{eq:Qhat estimate ss}) results in a biased estimate of $\mathbf{Q}$. Second, CM cannot accurately extrapolate $\mathbf{Q}$ over a gap in measurements. As can be seen in Eq. (\ref{eq:Q}), an abnormally long measurement interval $[t_{k-1}, t_k]$ due to a measurement outage will likely have a process noise covariance that is significantly different from previous shorter measurement intervals given the same fidelity filter dynamics model. Depending on the measurement type and mission scenario, measurement outages may be long and frequent as is often the case in angles-only relative navigation\cite{sullivan_angles_2018}. Third, CM techniques do not explicitly take into account that process noise is due to continuous-time unmodeled accelerations. Eq. (\ref{eq:Qhat estimate ss}) can result in any positive semi-definite $\mathbf{\hat{Q}}_{k}$, and there is no guarantee that a positive semi-definite $\mathbf{\widetilde{Q}}$ exists that relates to $\mathbf{\hat{Q}}_{k}$ through Eq. (\ref{eq:Q}). Lastly, to the authors' knowledge there has been no work on CM adaptive filtering for systems with colored process noise. These shortcomings are remedied by two novel techniques presented in the following section.

\section{Adaptive and Dynamically Constrained Process Noise Estimation}
\begin{figure}[!h]
\centering
\includegraphics[width=.8\linewidth,trim=170 540 256 100,clip]{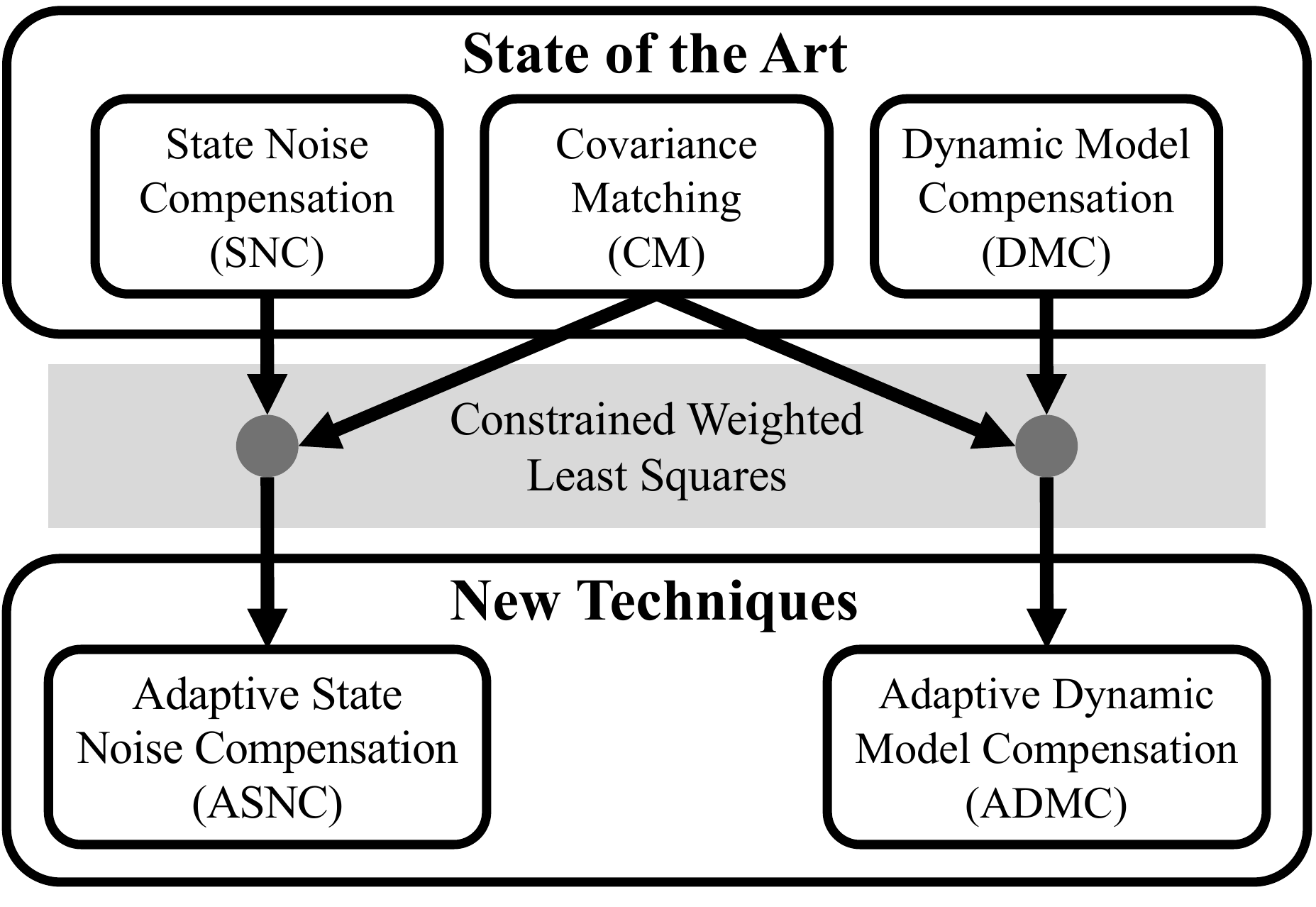}
\caption{Conceptual block diagram description of ASNC and ADMC.}
\label{fig:block diagram}
\end{figure}

This section describes how CM can be optimally combined with either SNC or DMC to overcome the limitations of the individual techniques. As illustrated in Fig. \ref{fig:block diagram}, this results in two adaptive and dynamically constrained process noise covariance estimation algorithms called ASNC and ADMC. Unlike SNC and DMC, the proposed techniques are adaptive and well suited for scenarios where the process noise covariance is time-varying and the dynamical environment is poorly known a priori. Furthermore, ASNC and ADMC are more accurate than CM because the first two terms in Eq. (\ref{eq:Qhat estimate}) are not neglected and the estimated process noise covariance is constrained according to the underlying continuous-time dynamical model. The proposed techniques also accurately extrapolate the process noise covariance over measurement outages and can easily leverage a priori bounds on the process noise power spectral density. ADMC enables better estimation accuracy than ASNC when the unmodeled accelerations are correlated in time provided that the measurement rate and accuracy are sufficient to enable accurate tracking of the unmodeled accelerations. 

The following two subsections describe how CM is optimally fused with SNC and DMC through a constrained weighted least squares minimization. In the next subsection, the weighting matrix in the least squares optimization is derived. The final subsection describes how the weighted least squares problem can be efficiently solved. Although SNC, DMC, CM, ASNC, and ADMC are each developed assuming a linear time-varying system, they can be applied in an EKF or a UKF to systems with a nonlinear dynamics model $\bm{x}_k = \bm{f}(\bm{x}_{k-1},\bm{u}_{k}) + \bm{w}_k$ and nonlinear measurement model $\bm{z}_k = \bm{h}(\bm{x}_k) + \bm{\nu}_k$. In this case, $\bm{\Delta}_k^z$ is the difference between the nonlinear true and predicted measurement vectors. In an EKF, the nonlinear models are linearized about the mean state estimate such that the state transition matrix and measurement matrix are replaced by the Jacobians $\mathbf{\Phi}_{k} = \frac{\partial \bm{f}(\bm{x}_{k-1},\bm{u}_k)}{\partial \bm{x}_{k-1}}\Bigr|_{\bm{\hat{x}}_{k-1|k-1}}$ and $\mathbf{H}_{k} = \frac{\partial \bm{h}(\bm{x}_k)}{\partial \bm{x}_k}\Bigr|_{\bm{\hat{x}}_{k|k-1}}$. For a UKF, Eq. (\ref{eq:Q}) utilizes the EKF linearization while other terms in the process noise algorithms containing $\mathbf{\Phi}$ or $\mathbf{H}$ are replaced by the corresponding UKF terms, which are computed using sigma points passed through $\bm{f}$ and $\bm{h}$.

\subsection{Adaptive State Noise Compensation}
Without loss of generality, the estimated state is written as
\begin{equation}\label{eq:generic state 1}
	\bm{x} = [\bm{x}_s^T \ \bm{p}^T]^T
\end{equation}
\noindent where $\bm{x}_s$ is a vector comprising the estimated spacecraft state, and $\bm{p}$ is a vector containing all other estimated parameters. Considering the state ordering in Eq. (\ref{eq:generic state 1}), the process noise covariance can be written in block matrix form as
\begin{equation}\label{eq:Q block ASNC}
	\mathbf{Q}
	=
	\begin{bmatrix}
		\mathbf{Q}_{ss} 		&\mathbf{Q}_{sp}\\
		\mathbf{Q}_{sp}^T 	&\mathbf{Q}_{pp}\\
	\end{bmatrix}
\end{equation}
where $\mathbf{Q}_{ss}$ is the process noise covariance of the spacecraft state, $\mathbf{Q}_{pp}$ is the process noise  covariance of the other estimated parameters, and $\mathbf{Q}_{sp}$ is the cross covariance. It is assumed that $\mathbf{\widetilde{Q}}(t)$ is constant over each measurement interval $[t_{k-1},t_k]$. Consequently, Eq. (\ref{eq:Q}) becomes
\begin{equation}\label{eq:Q const Qtilde}
	\mathbf{Q}_{k} = \int_{t_{k-1}}^{t_k}  \mathbf{\Phi}(t_k,\tau)  \mathbf{\Gamma}(\tau)  \mathbf{\widetilde{Q}}_{k}  \mathbf{\Gamma}(\tau)^T  \mathbf{\Phi}(t_k,\tau)^T  d\tau
\end{equation}
Due to the structure of Eq. (\ref{eq:Q const Qtilde}), $\mathbf{Q}_{k}$ is guaranteed to be positive semi-definite if $\mathbf{\widetilde{Q}}_{k} $ is positive semi-definite.

At each time step, ASNC finds the positive semi-definite $\mathbf{\widetilde{Q}}_{}$ that minimizes the difference in a weighted least squares sense between the elements of the spacecraft state process noise covariance $\mathbf{Q}_{ss}$ obtained through Eq. (\ref{eq:Q const Qtilde}) and the corresponding CM estimate obtained from Eq. (\ref{eq:Qhat estimate}). Since both the SNC modeled $\mathbf{Q}_{ss}$ and the corresponding CM estimate are symmetric, only the unique elements, which are contained in the lower triangular portions of the matrices, need to be fitted. The half-vectorization vech($\cdot$) indicates a vector composed of stacking the lower triangular elements of a matrix column-wise, which in this paper is also denoted by the superscript $vech$. For example, 
\begin{equation}
	\mathbf{M} = 
	\begin{bmatrix}
		M_{11}  \ M_{12}\\
		M_{21} \ M_{22}\\
	\end{bmatrix}
	\
	\Rightarrow
	\
	\text{vech}(\mathbf{M}) =\mathbf{M}^{vech} =
	\begin{bmatrix} 
	M_{11} \\ 
	M_{21} \\  
	M_{22}
	\end{bmatrix}
\end{equation}

In Eq. (\ref{eq:Q const Qtilde}), $\mathbf{Q}_{k}$ is a linear function of $\mathbf{\widetilde{Q}}_{k}$. To reduce computation time, $\mathbf{\widetilde{Q}}_{}$ is assumed diagonal in this work. Under this assumption, $\mathbf{Q}_{ss,k}^{vech}$ in Eq. (\ref{eq:Q const Qtilde}) can be expressed as
\begin{equation}\label{eq:Qunique func X}
	\mathbf{Q}_{ss,k}^{vech} = \mathbf{X}_{k}\mathbf{\widetilde{Q}}_{k}^{diag}
\end{equation}
where the vector $\mathbf{\widetilde{Q}}_{k}^{diag}$ is the main diagonal of $\mathbf{\widetilde{Q}}_{k}$. The matrix $\mathbf{X}_{k}$ is a function of $\bm{\Phi}$, $\bm{\Gamma}$, and the length of the measurement interval $[t_{k-1}, t_k]$. To construct $\mathbf{X}_{k}$, it is necessary to compute the integral in Eq. (\ref{eq:Q const Qtilde}), which can always be done numerically if $\bm{\Phi}$ and $\bm{\Gamma}$ are integrable functions. It is also straightforward to compute the integral analytically for linear time-invariant systems and some linear time-varying systems. At each time step, the adaptively tuned value of $\mathbf{\widetilde{Q}}^{diag}$ is found by solving the constrained weighted least squares minimization
\begin{align}\label{eq:ASNC Min}
&\underset{\mathbf{\widetilde{Q}}^{diag}}{\text{arg min}} \ 
	( \mathbf{X}_{k}\mathbf{\widetilde{Q}}^{diag} - \mathbf{\hat{Q}}_{ss,k+1}^{vech} )^T 
	\mathbf{W}_{k}^{-1} 
	( \mathbf{X}_{k}\mathbf{\widetilde{Q}}^{diag} - \mathbf{\hat{Q}}_{ss,k+1}^{vech} )\nonumber \\
&\text{subject to} \ \ \ \ \mathbf{\widetilde{Q}}_{l}^{diag} \leq \mathbf{\widetilde{Q}}^{diag} \leq \mathbf{\widetilde{Q}}_{u}^{diag}
\end{align}
Here $\mathbf{W}_{k}$ is the theoretical covariance of $\mathbf{\hat{Q}}_{ss,k+1}^{vech}$, which weights the solution of Eq. (\ref{eq:ASNC Min}) more heavily towards components of \smash{$\mathbf{\hat{Q}}_{ss,k+1}^{vech}$} that are known with more certainty. The inequality constraints in Eq. (\ref{eq:ASNC Min}) are element-wise where $\mathbf{\widetilde{Q}}_{l}^{diag}$ and $\mathbf{\widetilde{Q}}_{u}^{diag}$ specify lower and upper limits on $\mathbf{\widetilde{Q}}^{diag}$ based on coarse a priori knowledge of the dynamical environment. Each element of $\mathbf{\widetilde{Q}}_{l}^{diag}$ must be nonnegative to ensure the estimated $\mathbf{\widetilde{Q}}_{}$ is positive semi-definite. If no a priori knowledge of $\mathbf{\widetilde{Q}}_{}$ is available, the constraints imposed by $\mathbf{\widetilde{Q}}_{u}^{diag}$ can be omitted, and $\mathbf{\widetilde{Q}}_{l}^{diag}$ can be set to a vector of zeros. The matrix $\mathbf{\widetilde{Q}}_{k+1}$ is set equal to the diagonal $\mathbf{\widetilde{Q}}$ that solves the optimization in Eq. (\ref{eq:ASNC Min}). Then $\mathbf{Q}_{k+1}$ is computed through Eq. (\ref{eq:Q const Qtilde}), which adapts the process noise covariance according to the length of the measurement interval $[t_{k}, t_{k+1}]$. Like CM, the length of the sliding window is the only tunable parameter in ASNC. This parameter should be chosen based on how quickly $\mathbf{\widetilde{Q}}$ is expected to vary in time. 

\subsection{Adaptive Dynamic Model Compensation}
In ADMC, the state is augmented with empirical accelerations. Without loss of generality, the state and corresponding process noise covariance can be written as
\begin{equation}\label{eq:generic state DMC}
	\bm{x} = [\bm{x}_s^T \ \bm{\tilde{a}}^T \ \bm{p}^T]^T, \ \ \ \ \ 
		\mathbf{Q}
	=
	\begin{bmatrix}
		\mathbf{Q}_{ss} 					&\mathbf{Q}_{s\tilde{a}}					&\mathbf{Q}_{sp}\\
		\mathbf{Q}_{s\tilde{a}}^T 	&\mathbf{Q}_{\tilde{a}\tilde{a}}		&\mathbf{Q}_{\tilde{a}p}\\
		\mathbf{Q}_{sp}^T 				&\mathbf{Q}_{\tilde{a}p}^T				&\mathbf{Q}_{pp}
	\end{bmatrix}
\end{equation}
where $\mathbf{Q}_{\tilde{a}\tilde{a}}$ is the process noise covariance of the empirical accelerations. In ADMC, $\mathbf{\widetilde{Q}}$ is estimated  at each time step by solving the minimization in Eq. (\ref{eq:ASNC Min}). In this equation, $\mathbf{X}_{k}$ is different when using ADMC or ASNC since $\mathbf{\Phi}$ and $\mathbf{\Gamma}$ are altered when the state is augmented with empirical accelerations. Although $\mathbf{Q}_{s\tilde{a}}$ and $\mathbf{Q}_{\tilde{a}\tilde{a}}$ are linear functions of $\mathbf{\widetilde{Q}}$, simulations show it is detrimental to include CM estimates of these parameters in the optimization in Eq. (\ref{eq:ASNC Min}) to estimate $\mathbf{\widetilde{Q}}$. 

Similar to the CM approach in \cite{fraser_adaptive_2019}, a forgetting factor, $\alpha$, is introduced to smooth the estimate history of $\mathbf{\widetilde{Q}}$ in the presence of undesired oscillations. The filter can then be implemented as
\begin{equation}\label{eq:convex comb}
	\mathbf{\widetilde{Q}}_{k} = (1-\alpha)\mathbf{\widetilde{Q}}_{k-1}	 + \alpha\mathbf{\widetilde{Q}}_{k}^*	
\end{equation}
where $\mathbf{\widetilde{Q}}_{k}^*	$ is the diagonal matrix that minimizes Eq. (\ref{eq:ASNC Min}), and $0<\alpha\leq1$ is a selected constant. The value of $\alpha$ is a tunable parameter that provides greater smoothing for smaller values. The range $0.01\leq\alpha\leq0.05$ has been effective for the case studies presented in this paper and can be used as a guideline. After calculating $\mathbf{\widetilde{Q}}_{k}$ through Eq. (\ref{eq:convex comb}), Eq. (\ref{eq:Q const Qtilde}) is used to compute $\mathbf{Q}_{k+1}$ under the assumption that $\mathbf{\widetilde{Q}}_{k+1}=\mathbf{\widetilde{Q}}_{k}$. Thus, at each time step $k$ the developed algorithms assume $\mathbf{\widetilde{Q}}$ is constant over the interval $[t_{k-1}, t_{k+1}]$ because Eq. (\ref{eq:Q const Qtilde}) models $\mathbf{\widetilde{Q}}$ as constant over each measurement interval $[t_{k-1}, t_{k}]$, and it is assumed that $\mathbf{\widetilde{Q}}_{k+1} = \mathbf{\widetilde{Q}}_{k}$. It also assumed that $\mathbf{Q}$ is constant over the interval of the CM sliding window $[t_{k-N+1}, t_k]$ because the CM estimate in Eq. (\ref{eq:Qhat estimate}) is utilized in the optimization in Eq. (\ref{eq:ASNC Min}). Like ASNC, computing $\mathbf{Q}_{k+1}$ through Eq. (\ref{eq:Q const Qtilde}) adapts the process noise covariance according to the length of the measurement interval $[t_k, t_{k+1}]$. Note that the computed $\mathbf{Q}_{k+1}$ comprises the process noise covariance of the spacecraft state and the empirical accelerations. Estimating $\bm{\beta}$ as part of the filter state and eliminating the need for a tunable forgetting factor are left for future work.

\subsection{Weighting Matrix Derivation}
This section derives the least squares weighting matrix $\mathbf{W}_{k}$, which is the theoretical covariance matrix of $\mathbf{\hat{Q}}_{ss,{k+1}}^{vech}$. Specifically, the element in the $i$\textsuperscript{th} row and $j$\textsuperscript{th} column of $\mathbf{W}_{k}$ is 
\begin{equation}\label{eq:element W}
	W_{k_{i,j}} = \text{Cov}(\hat{Q}_{ss,{k+1_{i}}}^{vech}, \hat{Q}_{ss,{k+1_{j}}}^{vech})
\end{equation}
where $\hat{Q}_{ss,{k+1_{i}}}^{vech}$ is the $i$\textsuperscript{th} element of the vector $\mathbf{\hat{Q}}_{ss,{k+1}}^{vech}$. The sample variance of each element of $\mathbf{\hat{Q}}_{k+1}$ as shown in Eq. (\ref{eq:Qhat estimate}) is simply equal to the sample variance of each element of \smash{$\frac{1}{N}\sum_{p=k-N+1}^{k} \bm{\Delta}_{p}^x \bm{\Delta}_{p}^{x^T}$} since the other terms in the equation are considered deterministic. Both Kailath\cite{kailath_innovations_1968} and Mehra\cite{mehra_approaches_1972} provide proofs that the innovations, $\bm{\Delta}^z$, are a zero-mean white Gaussian process for an optimal Kalman filter at steady state. Therefore, it is apparent from Eq. (\ref{eq:state innovation}) that each $\bm{\Delta}_{k}^x$ is normally distributed because it is a linear transformation of a normally distributed random vector. Applying the expectation operator to Eq. (\ref{eq:state innovation}) also reveals that each $\bm{\Delta}_{k}^x$ is zero-mean. Furthermore, it is easily shown that $\bm{\Delta}_{}^x$ is uncorrelated in time by
\begin{align}
	\text{E}[\bm{\Delta}_{t}^x \bm{\Delta}_{\tau}^{x^T}] &= \mathbf{K}_{t}   \text{E}[\bm{\Delta}_t^z\bm{\Delta}_\tau^{z^T}]  \mathbf{K}_{\tau}^T\nonumber\\
																	&= \bm{0} \ \ \forall \ t\neq\tau
\end{align}
since $\text{E}[\bm{\Delta}_t^z \bm{\Delta}_\tau^{z^T}]= \bm{0} \ \ \forall \ t\neq\tau$\cite{kailath_innovations_1968,mehra_approaches_1972}. Note that Myers and Tapley state the assumption that $\bm{\Delta}_{}^x$ is not correlated in time in the original derivation of Eq. (\ref{eq:Qhat estimate})\cite{myers_adaptive_1976,myers_filtering_1974}. The theoretical covariance of $\bm{\Delta}_{k}^x$ is
\begin{align}
	\mathbf{\Sigma}_{k}   &=\text{E}[\bm{\Delta}_{k}^x \bm{\Delta}_{k}^{x^T}]\nonumber\\
																&= \mathbf{K}_{k} \mathbf{S}_{k}\mathbf{K}_{k}^T\label{eq:Sigma_k}
\end{align}

Now, two identities are employed. First, the covariance of two sums of random variables is given by
\begin{equation}\label{eq:cov sum}
	\text{Cov}\left(\sum_{i=1}^{n_i} X_i,\sum_{j=1}^{n_j} Y_j \right) = \sum_{i=1}^{n_i}\sum_{j=1}^{n_j}\text{Cov}(X_i,Y_j)
\end{equation}
where $X_i$ and $Y_j$ are random variables. Second, Isserlis' theorem states that for normally distributed random variables $X_1,...,X_4$,
\begin{align}\label{eq:isserlis}
	\text{E}[X_1X_2X_3X_4] = &\text{E}[X_1X_2]\text{E}[X_3X_4]+\text{E}[X_1X_3]\text{E}[X_2X_4]\nonumber \\
	&+\text{E}[X_1X_4]\text{E}[X_2X_3]
\end{align}
Using these two identities, the theoretical covariance of any two elements of the matrix $\mathbf{\hat{Q}}_{k+1} $ is given by
\begingroup
\allowdisplaybreaks
\begin{align}
	&\text{Cov}(\hat{Q}_{k+1_{a,b}} , \hat{Q}_{k+1_{m,n}}) \nonumber \\
										&= \text{Cov}\left(\frac{1}{N}\sum_{p=k-N+1}^{k} \Delta_{p_a}^x \Delta_{p_b}^x , \frac{1}{N}\sum_{p=k-N+1}^{k} \Delta_{p_m}^x \Delta_{p_n}^x  \right)\\
										&=\frac{1}{N^2}\text{Cov}\left(\sum_{p=k-N+1}^{k} \Delta_{p_a}^x \Delta_{p_b}^x , \sum_{p=k-N+1}^{k} \Delta_{p_m}^x \Delta_{p_n}^x  \right)\label{eq:Qhat cov 2}\\						
										&=\frac{1}{N^2}\sum_{p=k-N+1}^{k} \text{Cov}(\Delta_{p_a}^x \Delta_{p_b}^x, \Delta_{p_m}^x \Delta_{p_n}^x  )\label{eq:Qhat cov 3}\\
										&=\frac{1}{N^2}\sum_{p=k-N+1}^{k} (\text{E}[\Delta_{p_a}^x \Delta_{p_b}^x \Delta_{p_m}^x \Delta_{p_n}^x] \nonumber \\
										&\hspace{3.7cm} - \text{E}[\Delta_{p_a}^x \Delta_{p_b}^x] \text{E}[\Delta_{p_m}^x \Delta_{p_n}^x])\label{eq:Qhat cov 4}\\
										&=\frac{1}{N^2}\sum_{p=k-N+1}^{k} (\Sigma_{p_{a,m}} \Sigma_{p_{b,n}} + \Sigma_{p_{a,n}} \Sigma_{p_{b,m}} )\label{eq:Qhat cov 5}	
\end{align}
\endgroup
where the element in the $i$\textsuperscript{th} row and $j$\textsuperscript{th} column of $\mathbf{\hat{Q}}_{k+1}$ and $\mathbf{\Sigma}_{k}$ are given by $\hat{Q}_{k+1_{i,j}}$ and $\Sigma_{k_{i,j}}$ respectively. The $i$\textsuperscript{th} element of $\bm{\Delta}_k^x$ is $\bm{\Delta}_{k_i}^x$. Applying Eq. (\ref{eq:cov sum}) to Eq. (\ref{eq:Qhat cov 2}) and recalling the assumption that $\bm{\Delta}^x$ is uncorrelated in time yields Eq. (\ref{eq:Qhat cov 3}). The identity in Eq. (\ref{eq:isserlis}) can be applied to Eq. (\ref{eq:Qhat cov 4}) to yield Eq. (\ref{eq:Qhat cov 5}). In Eq. (\ref{eq:element W}), each $\hat{Q}_{ss,{k+1_{i}}}^{vech}$ corresponds to some $\hat{Q}_{k+1_{a,b}}$ in the lower triangular portion of $\mathbf{\hat{Q}}_{k+1}$. Thus each element of the weighting matrix $W_{k_{i,j}}$ can be computed through Eq. (\ref{eq:Qhat cov 5}) by determining the indices $a, \ b$ corresponding to $i$ and the indices $m, \ n$ corresponding to $j$.

In general, $\mathbf{W}_{k}$ is not a diagonal matrix. However, approximating $\mathbf{W}_{k}$ as diagonal by setting all off-diagonal elements equal to zero has multiple benefits. It guarantees that $\mathbf{W}_{k}$ is full rank, which ensures the optimization has a unique solution.  As described in the following subsection, a diagonal $\mathbf{W}_{k}$ also enables a more computationally efficient solution of the optimization. Computing the full weighting matrix using Eq. (\ref{eq:Qhat cov 5}) is cumbersome, but the diagonal approximation of $\mathbf{W}_{k}$ can be conveniently expressed as
\begin{equation}\label{eq:weighting matrix}
	\mathbf{W}_{k} = \frac{1}{N^2}\sum_{p=k-N+1}^{k}\bm{\mathcal{W}}_{p}
\end{equation}
where
\begin{align}
	\bm{\mathcal{W}}_{k} &= \text{diag}(\text{vech}(\mathbf{\overline{\Sigma}}_{k}))\\
	\mathbf{\overline{\Sigma}}_k &= \mathbf{\Sigma}_{ss,k}^{\circ 2}+\mathbf{\Sigma}_{ss,k}^{diag} \mathbf{\Sigma}_{ss,k}^{diag^T} \label{eq:sigma bar}
\end{align}
Here, diag($\cdot$) denotes a square diagonal matrix whose main diagonal is the vector inside the parenthesis. The matrix $\mathbf{\Sigma}_{ss,k}$ is the submatrix of $\mathbf{\Sigma}_k$ corresponding to the spacecraft state. For a six dimensional spacecraft state and the state orderings shown in Eqs. (\ref{eq:generic state 1}) and (\ref{eq:generic state DMC}), $\mathbf{\Sigma}_{ss,k}\in \mathbb{R}^{6\times6}$ is the first six rows and columns of $\mathbf{\Sigma}_k$. The vector $\mathbf{\Sigma}_{ss,k}^{diag}$ is the main diagonal of $\mathbf{\Sigma}_{ss,k}$. The Hadamard power, $^\circ$, denotes an element-wise power. For example, $\mathbf{A} = \mathbf{B}^{\circ 2}$ indicates that $A_{ij} = B_{ij}^2$. If the filter is at steady state such that ${\mathbf{\Sigma}}_{p} = {\mathbf{\Sigma}}_{k} \ \forall \ k-N< p \leq k$, Eq. (\ref{eq:weighting matrix}) reduces to
\begin{equation}\label{eq:weighting matrix ss}
		\mathbf{W}_{k} = \frac{1}{N} \bm{\mathcal{W}}_{k}
\end{equation}
The factor of $\frac{1}{N^2}$ in Eqs. (\ref{eq:Qhat cov 5}) and (\ref{eq:weighting matrix}) and the factor of $\frac{1}{N}$ in Eq. (\ref{eq:weighting matrix ss}) can be dropped since a constant factor does not change the solution of the least squares minimization in Eq. (\ref{eq:ASNC Min}).

\subsection{Constrained Weighted Least Squares Solution}
The ASNC and ADMC algorithms are summarized in Table \ref{tab:ASNC and ADMC algorithms}. On line four of Table \ref{tab:ASNC and ADMC algorithms}, $\mathbf{\widetilde{Q}}$ is estimated through a constrained weighted least squares optimization, which approximates the maximum likelihood estimate. Given a linear system, as shown in Eqs. (\ref{eq:true dynamics}) and (\ref{eq:linear measurement model}), and assuming an optimal Kalman filter with constant $\mathbf{Q}$, Eq. (\ref{eq:Qhat estimate}) provides an unbiased estimate of the process noise covariance\cite{myers_filtering_1974,myers_adaptive_1976}. According to the Central Limit Theorem, the probability distribution of each element of the last term in Eq. (\ref{eq:Qhat estimate}) approaches a normal distribution as the length of the sliding window increases. Therefore, the weighted least squares solution obtained through Eq. (\ref{eq:ASNC Min}) approaches the maximum likelihood estimate of $\mathbf{\widetilde{Q}}$ as the length of the sliding window increases. Since $\mathbf{Q}$ is likely to change significantly for an abnormally long measurement interval (see Eq. (\ref{eq:Q const Qtilde})), measurement intervals corresponding to a measurement outage should be excluded from the sliding windows in Eqs. (\ref{eq:Qhat estimate}) and (\ref{eq:weighting matrix}) to avoid biasing the estimate of $\mathbf{\widetilde{Q}}$. 

The optimization in Eq. (\ref{eq:ASNC Min}) is a quadratic program with linear inequality constraints. In general, this optimization can be efficiently solved using active set or interior point methods\cite{lawson_solving_1995,paige_computer_1979,paige_fast_1979,bjorck_numerical_1996}, and applicable solvers are readily available in various programming languages. However, Eq. (\ref{eq:ASNC Min}) can be solved with a significant reduction in computation time when $\mathbf{W}_{k}$ is approximated as diagonal, the spacecraft state is represented by Cartesian coordinates, and the continuous-time process noise $\bm{\epsilon}$ is expressed in the same frame as the spacecraft state.
 Under these conditions, each element of $\mathbf{\widetilde{Q}}^{diag^*}_{k}$ can be solved for independently without any matrix inversions or decompositions. This computationally efficient solution is described in the remainder of this subsection for both ASNC and ADMC.

\begin{table}[!t]
\renewcommand{\arraystretch}{1.3} 
\caption{Summary of the ASNC and ADMC algorithms.}
\label{tab:ASNC and ADMC algorithms}
\centering
\begin{tabular}{l l}
\toprule
\toprule
1:		&Compute $\mathbf{\hat{Q}}_{ss,{k+1}}$ using Eq. (\ref{eq:Qhat estimate})\\
2:		&Calculate $\mathbf{\Sigma}_{k}$ through Eq. (\ref{eq:Sigma_k})\\
3:		&Compute $\mathbf{W}_{k}$ using Eqs. (\ref{eq:weighting matrix}-\ref{eq:sigma bar})\\
4: 	&Determine $\mathbf{\widetilde{Q}}^{*}_{k}$ by solving the optimization in Eq. (\ref{eq:ASNC Min}) \\
		&Solve Eq. (\ref{eq:ASNC Min}) using Eqs. (\ref{eq:min efficient vecs}-\ref{eq:sol ASNC}) if applicable\\
5:		&Calculate $\mathbf{\widetilde{Q}}_{k}$ through Eq. (\ref{eq:convex comb}) ($\alpha = 1$ for ASNC)\\
6: 	&Set $\mathbf{\widetilde{Q}}_{k+1}=\mathbf{\widetilde{Q}}_{k}$\\
7: 	&Compute $\mathbf{Q}_{k+1}$ using Eq. (\ref{eq:Q const Qtilde})\\
\bottomrule
\bottomrule
\end{tabular}
\end{table}

\subsubsection{Adaptive State Noise Compensation.}
Consider the state shown in Eq. (\ref{eq:generic state 1}) where the spacecraft state is
\begin{equation}\label{eq:Cart sc state}
	\bm{x}_s = [\bm{r}^T \ \bm{v}^T]^T
\end{equation}
Here $\bm{r} \in \mathbb{R}^3$ and $\bm{v} \in \mathbb{R}^3$ are the spacecraft position and velocity vectors respectively expressed in an inertial frame. Recall the corresponding process noise covariance in block matrix form shown in Eq. (\ref{eq:Q block ASNC}). To avoid the computational cost of numerically evaluating Eq. (\ref{eq:Q const Qtilde}), an analytical solution derived by Myers\cite{myers_filtering_1974} can be utilized which is given by
\begin{equation}\label{eq:Q ASNC efficient}
\mathbf{Q}_{ss,k} = 
    \begin{bmatrix}
        \frac{1}{3}\Delta t_k^3 \mathbf{\widetilde{Q}}_{k}  &\frac{1}{2}\Delta t_k^2 \mathbf{\widetilde{Q}}_{k}\\
        \frac{1}{2}\Delta t_k^2\mathbf{\widetilde{Q}}_{k}   &\Delta t_k\mathbf{\widetilde{Q}}_{k}
    \end{bmatrix}
\end{equation}
when the continuous-time process noise $\bm{\epsilon}$ is expressed in the same frame as the spacecraft state. Here $\Delta t_k = t_k - t_{k-1}$ is the length of the measurement interval. As detailed in \cite{carpenter_navigation_2018} and \cite{shalom_estimation_2002}, Eq. (\ref{eq:Q ASNC efficient}) is an approximate solution to Eq. (\ref{eq:Q const Qtilde}) that assumes $\Delta t_k$ is small. The size of $\Delta t_k$ for which Eq. (\ref{eq:Q ASNC efficient}) can be accurately applied is characterized in \cite{stacey_process_2021}. Additional analytical approximations of Eq. (\ref{eq:Q const Qtilde}) for absolute and relative spacecraft states parameterized in both Cartesian coordinates and orbital elements are provided in \cite{stacey_process_2021}.

Notice that in Eq. (\ref{eq:Q ASNC efficient}) each element of $\mathbf{Q}_{ss}$ is a function of a single element of $\mathbf{\widetilde{Q}}$. Approximating $\mathbf{W}_{k}$ as diagonal through Eq. (\ref{eq:weighting matrix}), the optimization in Eq. (\ref{eq:ASNC Min}) can be written as
\begin{align}\label{eq:ASNC Min Efficient}
 &\underset{\mathbf{\widetilde{Q}}^{diag}}{\text{arg min}} \sum_{i=1}^3
	( \bm{\overline{X}}(i)\widetilde{Q}_{i}^{diag} - \bm{b}(i) )^T 
	\mathbf{\overline{W}}(i)^{-1} 
	( \bm{\overline{X}}(i)\widetilde{Q}_{i}^{diag} - \bm{b}(i) )\nonumber \\
	&\text{subject to} \ \ \ \ \mathbf{\widetilde{Q}}_{l}^{diag} \leq\mathbf{\widetilde{Q}}_{}^{diag} \leq \mathbf{\widetilde{Q}}_{u}^{diag}
\end{align}
where $\widetilde{Q}_{i}^{diag}$ is the $i$\textsuperscript{th} element of $\mathbf{\widetilde{Q}}^{diag}$ and
\begin{align}\label{eq:min efficient vecs}
	&\bm{\overline{X}}(i) = 
		\begin{bmatrix}
			\frac{1}{3}\Delta t_k^3\\
			\frac{1}{2}\Delta t_k^2\\
			\Delta t_k
		\end{bmatrix}
		,\hspace{0.5cm}
		\bm{b}(i) = 
			\begin{bmatrix}
				\hat{Q}_{ss,k+1_{i,i}}\\
				\hat{Q}_{ss,k+1_{3+i,i}}\\
				\hat{Q}_{ss,k+1_{3+i,3+i}}
			\end{bmatrix}
		,\nonumber \\
		&\mathbf{\overline{W}}(i) = \text{diag}\left(
		 \sum_{p=k-N+1}^{k}
			\begin{bmatrix}
				\overline{\Sigma}_{p_{i,i}}\\
				\overline{\Sigma}_{p_{3+i,i}}\\
				\overline{\Sigma}_{p_{3+i,3+i}}
			\end{bmatrix}\right)
\end{align}
The objective function in Eq. (\ref{eq:ASNC Min Efficient}) is the sum of three independent quadratic functions which are each a function of a single optimization variable. Consequently, each element of the solution to Eq. (\ref{eq:ASNC Min Efficient}), $\mathbf{\widetilde{Q}}^{diag^*}_{k}$, can be solved for independently. The solution to the optimization in Eq. (\ref{eq:ASNC Min Efficient}) can be written as
\begin{align} 
\bar{Q}_{i}^* &= \frac{\overline{\bm{X}}(i)^T\mathbf{\overline{W}}(i)^{-1}\bm{b}(i)}{\overline{\bm{X}}(i)^T\mathbf{\overline{W}}(i)^{-1}\overline{\bm{X}}(i)}\label{eq:unconstrained solution}\\[4pt]
\widetilde{Q}^{diag^*}_{i}  &= 
     \begin{cases}
       \widetilde{Q}_{l_i}^{diag} &\ \ \text{if} \ \ \ \bar{Q}_{i}^*< \widetilde{Q}_{l_i}^{diag}\\
       \widetilde{Q}_{u_i}^{diag} &\ \ \text{if} \ \ \ \bar{Q}_{i}^*> \widetilde{Q}_{u_i}^{diag} \\
        \bar{Q}_{i}^* &\ \ \text{otherwise}\\
     \end{cases}\label{eq:sol ASNC}
\end{align} 
where $\bm{\bar{Q}}^*$ is the solution to Eq. (\ref{eq:ASNC Min Efficient}) if the constraints are removed, and the subscript $i$ denotes the $i$\textsuperscript{th} component of a vector. If no upper limit on $\mathbf{\widetilde{Q}}^{diag}$ is included in Eq. (\ref{eq:ASNC Min Efficient}), the second conditional of Eq. (\ref{eq:sol ASNC}) is omitted. Since $\mathbf{\overline{W}}$ is diagonal, its inverse in Eq. (\ref{eq:unconstrained solution}) can be computed element-wise on the diagonal elements. Note that Eqs. (\ref{eq:unconstrained solution}-\ref{eq:sol ASNC}) are an exact solution of the optimization shown in Eq. (\ref{eq:ASNC Min Efficient}). This solution is computationally efficient because it does not require any matrix inversions or matrix decompositions.

\subsubsection{Adaptive Dynamic Model Compensation.} 
Consider the state and corresponding block matrix process noise covariance in Eq. (\ref{eq:generic state DMC}) where the spacecraft state is given by Eq. (\ref{eq:Cart sc state}).  In this case an analytical approximation of Eq. (\ref{eq:Q const Qtilde}) was derived by Cruickshank\cite{cruickshank_genetic_1998}, which is
\begingroup
\allowdisplaybreaks
\begin{align}\label{eq:Q ADMC efficient}
&\begin{bmatrix}
\mathbf{Q}_{ss,k}  				&\mathbf{Q}_{s\tilde{a},k}\\
\mathbf{Q}_{s\tilde{a},k}^T	&\mathbf{Q}_{\tilde{a} \tilde{a},k}
\end{bmatrix}
 = 
    \begin{bmatrix}
        \mathbf{C}_{1,1}\circ \mathbf{\widetilde{Q}}_{k}  &\mathbf{C}_{2,1}\circ\mathbf{\widetilde{Q}}_{k}  &\mathbf{C}_{3,1}\circ\mathbf{\widetilde{Q}}_{k}\\
        \mathbf{C}_{2,1}\circ\mathbf{\widetilde{Q}}_{k}   &\mathbf{C}_{2,2}\circ\mathbf{\widetilde{Q}}_{k}  &\mathbf{C}_{3,2}\circ\mathbf{\widetilde{Q}}_{k}\\
        \mathbf{C}_{3,1}\circ\mathbf{\widetilde{Q}}_{k}   &\mathbf{C}_{3,2}\circ\mathbf{\widetilde{Q}}_{k}  &\mathbf{C}_{3,3}\circ\mathbf{\widetilde{Q}}_{k}\\
    \end{bmatrix}\nonumber\\
    &C_{1,1_{i}}  = \frac{\Delta t_k^3}{3\beta_{i}^2} - \frac{\Delta t_k^2}{\beta_{i}^3} + \frac{\Delta t_k}{\beta_{i}^4}(1-2e^{-\beta_{i}\Delta t_k}) \nonumber \\
    &\hspace{1.2cm}+ \frac{1}{2\beta_{i}^5}(1-e^{-2\beta_{i}\Delta t_k})\nonumber \\
	&C_{2,1_{i}}  = \frac{\Delta t_k^2}{2\beta_{i}^2} - \frac{\Delta t_k}{\beta_{i}^3}(1-e^{-\beta_{i}\Delta t_k}) + \frac{1}{\beta_{i}^4}(1-e^{-\beta_{i}\Delta t_k}) \nonumber \\
	&\hspace{1.2cm}- \frac{1}{2\beta_{i}^4}(1-e^{-2\beta_{i}\Delta t_k})\nonumber \\
	&C_{3,1_{i}}  = \frac{1}{2\beta_{i}^3}(1-e^{-2\beta_{i}\Delta t_k}) - \frac{\Delta t_k}{\beta_{i}^2} e^{-\beta_{i}\Delta t_k}\nonumber \\
	&C_{2,2_{i}}  = \frac{\Delta t_k}{\beta_{i}^2} - \frac{2}{\beta_{i}^3}(1-e^{-\beta_{i}\Delta t_k}) + \frac{1}{2\beta_{i}^3}(1-e^{-2\beta_{i}\Delta t_k})\nonumber \\
	&C_{3,2_{i}}  = \frac{1}{2\beta_{i}^2}(1+e^{-2\beta_{i}\Delta t_k}) - \frac{1}{\beta_{i}^2}e^{-\beta_{i}\Delta t_k}\nonumber \\
	&C_{3,3_{i}}  = \frac{1}{2\beta_{i}}(1-e^{-2\beta_{i}\Delta t_k})
\end{align}
\endgroup
when $\bm{\beta}$ and $\mathbf{\widetilde{Q}}$ are diagonal matrices, and the continuous-time process noise $\bm{\epsilon}$ is expressed in the same frame as the spacecraft state\cite{cruickshank_genetic_1998,goldstein_real-time_2000,schutz_statistical_2004}. As explained in \cite{carpenter_navigation_2018}, Eq. (\ref{eq:Q ADMC efficient}) assumes $\Delta t_k$ is small. Here $\mathbf{C}_{1,1}, \hdots, \mathbf{C}_{3,3}\in\mathbb{R}^{3\times3}$ are each a diagonal matrix, and $C_{1,1_i}$ and $\beta_i$ refer to the $i$\textsuperscript{th} diagonal element of the matrices $\mathbf{C}_{1,1}$ and $\bm{\beta}$ respectively. The Hadamard product $\circ$ denotes element-wise multiplication. For example, $\mathbf{A} = \mathbf{B}\circ\mathbf{C}$ indicates that $A_{ij} = B_{ij}C_{ij}$. Similar to ASNC, each element of $\mathbf{Q}_{ss}$ is a function of only a single element of $\mathbf{\widetilde{Q}}$. Approximating $\mathbf{W}$ as diagonal, the optimization in Eq. (\ref{eq:ASNC Min}) can be rewritten as Eq. (\ref{eq:ASNC Min Efficient}) where $\bm{b}(i)$ and $\mathbf{\overline{W}}(i)$ are defined in Eq. (\ref{eq:min efficient vecs}) and
\begin{equation}
	\bm{\overline{X}}(i) = 
		\left[C_{1,1_i}  \ \ C_{2,1_i} \ \  C_{2,2_i}    \right]^T
\end{equation}
The solution to Eq. (\ref{eq:ASNC Min Efficient}) is given by Eqs. (\ref{eq:unconstrained solution}) and (\ref{eq:sol ASNC}).

\section{Case Study I - Particle in One Dimension}
This section compares the developed algorithms with several state of the art process noise techniques through a linear system. A similar example is used by several authors to delineate SNC and DMC\cite{schutz_statistical_2004,goldstein_real-time_2000,cruickshank_genetic_1998}. Consider a particle moving in one dimension. Range and range-rate measurements are taken from the origin to the particle every 0.1 s and are corrupted by zero-mean white Gaussian noise with standard deviations of 2 m and 0.1 m/s respectively. A Kalman filter estimates the particle position and velocity over 240 s in two scenarios where the particle is subject to an unknown perturbing acceleration $a_p(t)$. In the first scenario, $a_p(t)$ is a zero-mean white Gaussian process where $\text{E}[a_p(t)a_p(\tau)] = \frac{1}{2}\delta(t-\tau)$ m$^2$/s$^4$. For the second scenario, $a_p(t) = \text{cos}\left(\frac{\pi}{5}t\right) \ \text{m/s\textsuperscript{2}}$ is a deterministic unmodeled acceleration, which is more similar to what is typically encountered in orbit determination. 

CM as described by Eq. (\ref{eq:Qhat estimate ss}) as well as ASNC and ADMC are each utilized with a sliding window of 30 measurement intervals, which is 3 s. Adaptation of the process noise covariance does not commence until the 31\textsuperscript{st} filter call when the entire sliding window is filled. For ASNC and ADMC, no upper bound on $\mathbf{\widetilde{Q}}^{diag}$ is utilized, and the lower bound is set to a vector of zeros. A promising new process noise covariance estimation technique developed by Moghe et al.\cite{moghe_adaptive_2019} is also simulated. This approach is restricted to linear time-invariant systems and is guaranteed to converge to the true $\mathbf{Q}$ in the stochastic acceleration scenario. Although the method estimates $\mathbf{Q}$ after each filter call, the estimate is not guaranteed positive definite. If an estimate of $\mathbf{Q}$ is not positive definite, it is set to the previous estimate, and the filter $\mathbf{Q}$ is not updated. Additionally, the Bayesian interacting multiple model (IMM) approach presented by Li and Bar-Shalom\cite{li_recursive_1994} is simulated. This method requires two filters to run in parallel, each modeling a different system mode. In this case, the first filter utilizes a lower bound on the process noise power spectral density $\widetilde{Q}_{min} = 10^{-3}$ m$^2$/s$^{3}$ and the second filter utilizes an upper bound $\widetilde{Q}_{max} = 100$ m$^2$/s$^{3}$. The stochastic acceleration scenario is also simulated with $\widetilde{Q}_{min} = 10^{-2}$ m$^2$/s$^{3}$ and $\widetilde{Q}_{max} = 1$ m$^2$/s$^{3}$. Eq. (\ref{eq:Q const Qtilde}) is used to compute $\mathbf{Q}$ from $\widetilde{Q}$. The combined estimate of $\mathbf{Q}$ at time step $k$ is computed in terms of its matrix square root $\mathbf{Q}_k = \mathbf{Q}^{1/2}_k(\mathbf{Q}^{1/2}_k)^T$ where $\mathbf{Q}^{1/2}_k = \sum_{i=1}^2 (\mathbf{Q}^i)^\frac{1}{2} \mu^i_k$. Here $(\mathbf{Q}^i)^\frac{1}{2}$ is the matrix square root of the process noise covariance used in the $i$\textsuperscript{th} filter. The probability that the $i$\textsuperscript{th} mode is active after processing all measurements through time step $t_k$ is denoted by $\mu^i_k$. The initial mode probabilities are set to provide the combined $\mathbf{Q}$ corresponding to the specified initial $\widetilde{Q}$. The matrix $\bm{\pi}\in \mathbb{R}^{2\times2}$ is fixed where $\pi_{i,j}$ denotes the probability of transitioning from mode $i$ to mode $j$. For these simulations,
\begin{equation}\label{eq:transition probabilities}
	\bm{\pi} = 
	\begin{bmatrix}
			0.99 	&0.01\\
			0.01	    &0.99
	\end{bmatrix}
\end{equation}

The performance of each considered process noise technique is characterized through Monte Carlo (MC) simulations where the initial 1-$\sigma$ formal uncertainties provided to the filter in position and velocity are 1.8 m and 150 mm/s. In each simulation, the initial estimate error in position and velocity is randomly sampled from a zero-mean normal distribution whose covariance is the initial formal covariance provided to the filter. For DMC and ADMC, the empirical acceleration is initialized as zero. To reduce the memory required by ASNC and ADMC, Eq. (\ref{eq:weighting matrix ss}) is used to approximate the weighting matrix. This approximation provides nearly identical performance to computing the weighting matrix through Eq. (\ref{eq:weighting matrix}).

\subsection{Adaptive State Noise Compensation}
The estimated state is
\begin{equation}
	\bm{x} = 
	[ x \ \ \dot{x}]^T	
\end{equation} 
where $x$ and $\dot{x}$ are the particle position and velocity respectively. The plant matrix,  process noise mapping matrix, state transition matrix, and measurement matrix are
\begin{alignat}{2}
	\mathbf{A} &=
	\begin{bmatrix}
		0 	&1\\
		0 	&0
	\end{bmatrix}
	,\hspace{.4cm}
	&&\mathbf{\Gamma} = 
	\begin{bmatrix}
		0\\
		1
	\end{bmatrix}
	,\nonumber \\
	\mathbf{\Phi}(t_k,t_{k-1}) &= 
	\begin{bmatrix}
		1 	&\Delta t_k \\
		0 	&1
	\end{bmatrix}
	,\hspace{.4cm}
	&&\mathbf{H} = 
	\begin{bmatrix}
		1 	&0\\
		0 	&1
	\end{bmatrix}
\end{alignat}
Since this is a linear time-invariant system, it is straightforward to analytically evaluate the integral in Eq. (\ref{eq:Q const Qtilde}) which yields
\begin{equation}\label{eq:Q SNC Case I}
	\mathbf{Q}_{k} = \widetilde{Q}_k
	\begin{bmatrix}
		\frac{1}{3}\Delta t_k^3 	&\frac{1}{2}\Delta t_k^2\\
		\frac{1}{2}\Delta t_k ^2	&\Delta t_k 
 	\end{bmatrix}
\end{equation}
For this one dimensional system, $\widetilde{Q}$ is a scalar. The diagonal weighting matrix is
\begingroup
\allowdisplaybreaks
\begin{flalign}
	&\mathbf{W}_k = \text{diag}([\text{Var}(\hat{Q}_{ss,{k+1_{1}}}^{vech}) \ \text{Var}(\hat{Q}_{ss,{k+1_{2}}}^{vech}) \ \text{Var}(\hat{Q}_{ss,{k+1_{3}}}^{vech})]^T)\label{eq:example W elements}\\
						 &=  \text{diag}([\text{Var}(\hat{Q}_{{k+1_{1,1}}}) \ \text{Var}(\hat{Q}_{{k+1_{2,1}}}) \ \text{Var}(\hat{Q}_{{k+1_{2,2}}})]^T)\label{eq:example W Qhat}\\
						 &= \frac{1}{N^2} \sum_{p=k-N+1}^k \hspace{-.1cm}\text{diag}([2\Sigma_{p_{1,1}}^2 \ \ \Sigma_{p_{2,1}}^2 + \Sigma_{p_{1,1}}\Sigma_{p_{2,2}} \ \ 2\Sigma_{p_{2,2}}^2]^T)\label{eq:example W Sigma}\\
						 &\approx \frac{1}{N} \text{diag}([2\Sigma_{k_{1,1}}^2 \ \ \Sigma_{k_{2,1}}^2 + \Sigma_{k_{1,1}}\Sigma_{k_{2,2}} \ \ 2\Sigma_{k_{2,2}}^2]^T)\label{eq:example W steady state}
\end{flalign}
\endgroup
where $\text{Var}(X) = \text{Cov}(X,X)$ is the variance of the random variable $X$. Applying Eq. (\ref{eq:element W}) to the diagonal elements of $\mathbf{W}_k$ results in Eq. (\ref{eq:example W elements}), which is equivalent to Eq. (\ref{eq:example W Qhat}) by definition of the half-vectorization. Applying Eqs. (\ref{eq:weighting matrix} - \ref{eq:sigma bar}) yields Eq. (\ref{eq:example W Sigma}). Note that Eq. (\ref{eq:example W Sigma}) can also be obtained by applying Eq. (\ref{eq:Qhat cov 5}) to Eq. (\ref{eq:example W Qhat}). For this case study, the approximation in Eq. (\ref{eq:weighting matrix ss}) is utilized, resulting in Eq. (\ref{eq:example W steady state}). ASNC adaptively tunes $\widetilde{Q}$ at each time step through Eqs. (\ref{eq:unconstrained solution}-\ref{eq:sol ASNC}). In Eq. (\ref{eq:unconstrained solution}), $\bm{b}(i) = [\hat{Q}_{ss,k+1_{1,1}} \ \ \hat{Q}_{ss,k+1_{2,1}} \ \ \hat{Q}_{ss,k+1_{2,2}}]^T$ and $\mathbf{\overline{W}}(i)$ is given by Eq. (\ref{eq:example W steady state}) because the system is one dimensional and the approximation in Eq. (\ref{eq:weighting matrix ss}) is utilized. The estimated value of $\widetilde{Q}$ is then used to compute $\mathbf{Q}_{k+1}$ through Eq. (\ref{eq:Q SNC Case I}). 

\subsection{Adaptive Dynamic Model Compensation}
For ADMC, the estimated state is augmented with an empirical acceleration. The new state vector is
\begin{equation}
	\bm{x} = [ x \ \ \dot{x} \ \ \tilde{a}]^T	
\end{equation} 
The plant matrix and process noise mapping matrix are
\begin{equation}
	\mathbf{A} = 
	\begin{bmatrix}
		0 	&1 	&0\\
		0 	&0 	&1\\
		0 	&0 	&-\beta
	\end{bmatrix}
	, \ \ \ \ 
	\mathbf{\Gamma} = 
	\begin{bmatrix}
		0\\
		0\\
		1
	\end{bmatrix}
\end{equation}
The parameters $\beta$ and $\alpha$ are set to 0.005 s\textsuperscript{-1} and 0.02 respectively for this case study. The state transition matrix and measurement sensitivity matrix are
\begin{align}
	\mathbf{\Phi}(t_k,t_{k-1}) &= 
	\begin{bmatrix}
		1 	&\Delta t_k 	&\frac{1}{\beta}\Delta t_k 	+ \frac{1}{\beta^2}(e^{-\beta			\Delta t_k}-1)\\
		0 	&1 			&\frac{1}{\beta}(1-e^{-\beta\Delta t_k})\\
		0 	&0 			&e^{-\beta\Delta t_k}
	\end{bmatrix}
	 \nonumber \\ 
	\mathbf{H} &= 
	\begin{bmatrix}
		1 	&0 	&0\\
		0 	&1 	&0
	\end{bmatrix}
\end{align}
\noindent Evaluating Eq. (\ref{eq:Q const Qtilde}) yields
\begin{equation}\label{eq:Q DMC Case I}
	\mathbf{Q}_{k} = \tilde{Q}_k
	\begin{bmatrix}
		C_{1,1} 	&C_{2,1} 	&C_{3,1}\\
		C_{2,1} 	&C_{2,2} 	&C_{3,2}\\
		C_{3,1} 	&C_{3,2} 	&C_{3,3}\\
	\end{bmatrix}\\
\end{equation}
Note that $C_{1,1},\hdots,C_{3,3}$ are scalars for this one dimensional system and are defined in Eq. (\ref{eq:Q ADMC efficient}). The weighting matrix is given by Eq. (\ref{eq:example W steady state}). ADMC adaptively tunes $\widetilde{Q}$ at each time step through Eqs. (\ref{eq:unconstrained solution}-\ref{eq:sol ASNC}). The solution to Eq. (\ref{eq:ASNC Min}), $\widetilde{Q}^*_k$, is used to compute $\widetilde{Q}_{k+1}$ through Eq. (\ref{eq:convex comb}), which is then used to calculate $\mathbf{Q}_{k+1}$ through Eq. (\ref{eq:Q DMC Case I}). The eigenvector decompositions of Eqs. (\ref{eq:Q SNC Case I}) and (\ref{eq:Q DMC Case I}) illustrate how the developed algorithms constrain the process noise covariance for this system according to the underlying continuous-time dynamical model. In Eqs. (\ref{eq:Q SNC Case I}) and (\ref{eq:Q DMC Case I}), the eigenvectors  and the ratio of eigenvalues of $\mathbf{Q}$ are fixed. The chosen value of $\widetilde{Q}$ only scales the magnitudes of the eigenvalues of $\mathbf{Q}$. Therefore, the shape and orientation of the 1-$\sigma$ uncertainty ellipsoid associated with $\mathbf{Q}$ are fixed, and the chosen value of $\widetilde{Q}$ only scales that ellipsoid.

\subsection{Results}
\begin{figure*}[!h]
\centering 
\subfigure[Position]{\label{ffig:MAE a}
\includegraphics[width=.27\linewidth,trim= 0 0 -2 0,clip]{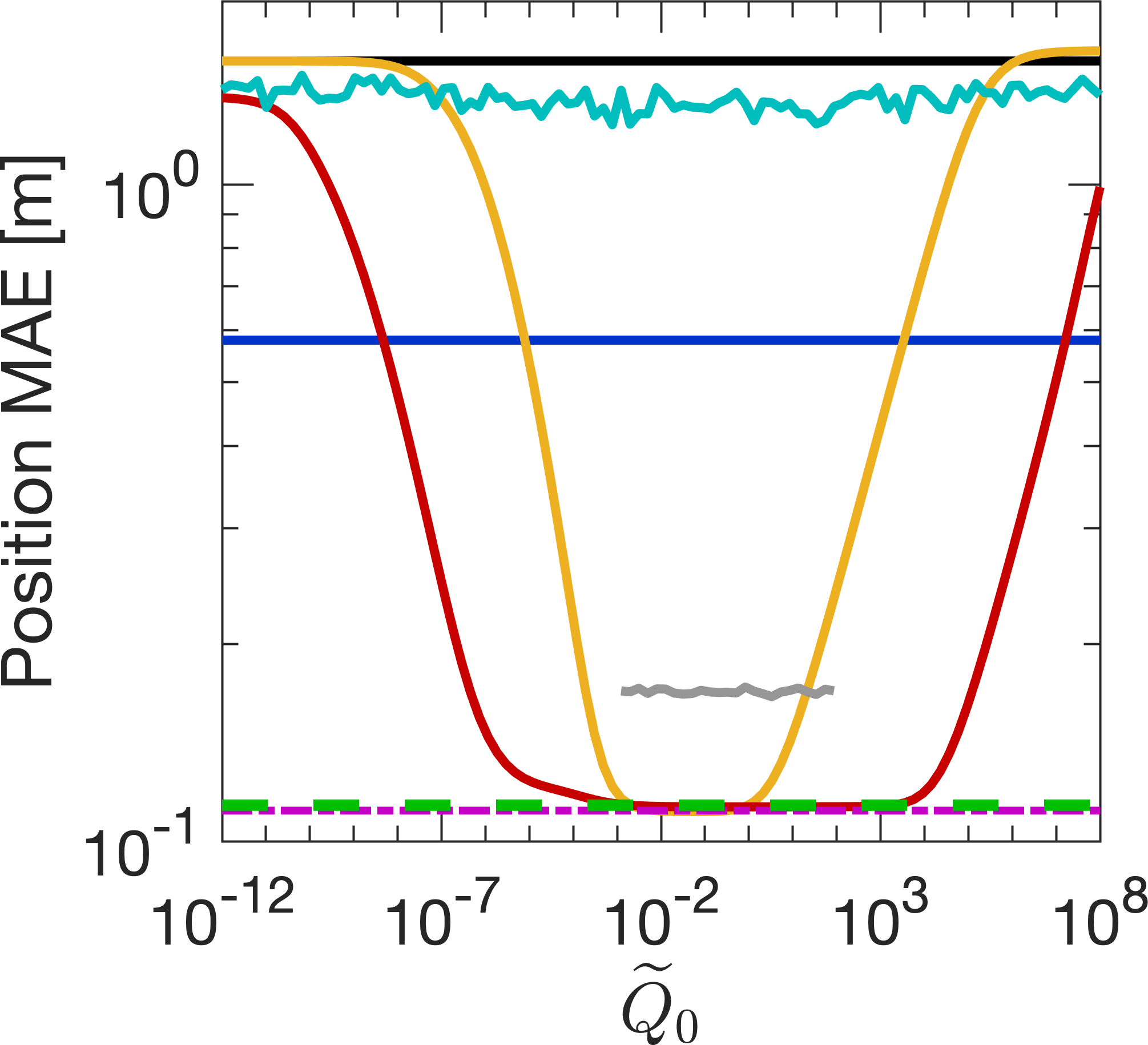}}
\subfigure[Velocity]{\label{fig:MAE b}
\includegraphics[width=.27\linewidth,trim= -1 0 -1 0,clip]{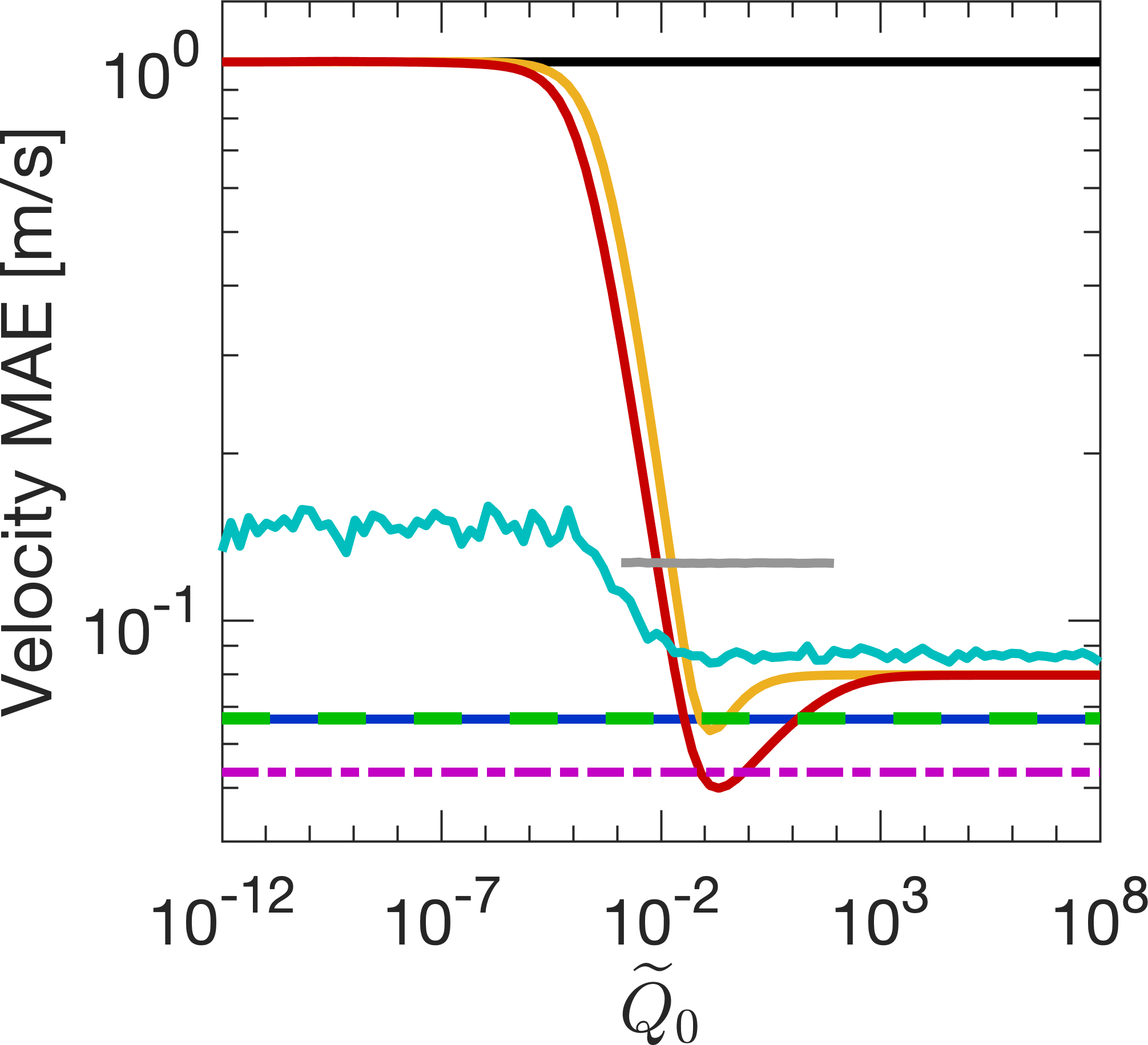}}
\subfigure[Acceleration]{\label{fig:MAE c}
\includegraphics[width=.27\linewidth,trim= -2 0 0 0,clip]{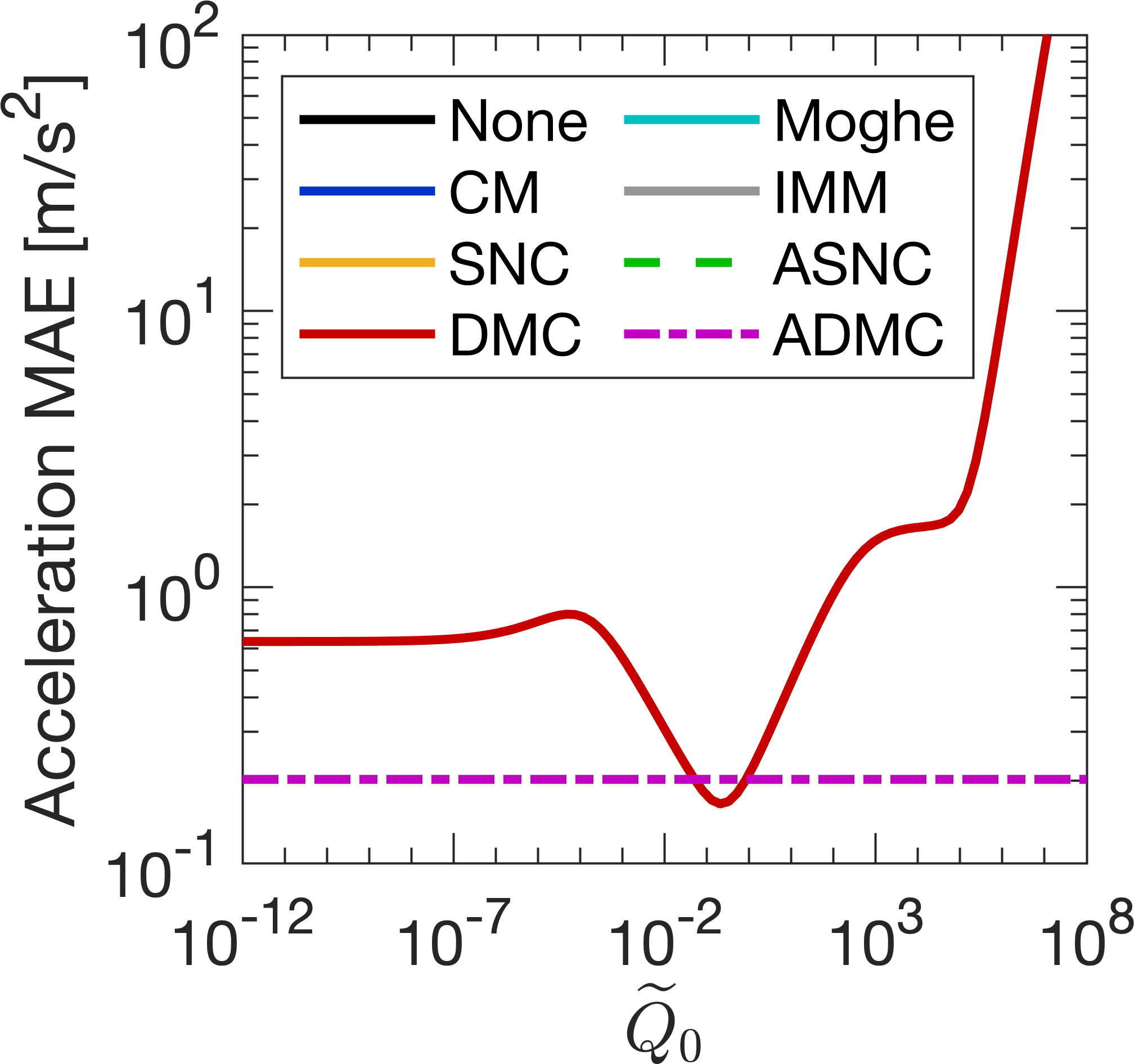}}
\caption{Case study I MAE for the deterministic acceleration scenario. The units of the initial guess of the process noise power spectral density, $\widetilde{Q}_0$, are m$^2$/s$^{3}$ for SNC and m$^2$/s$^{5}$ for DMC\cite{schutz_statistical_2004}.}
\label{fig:MAE}
\end{figure*}

The estimation mean absolute error (MAE) for the stochastic and deterministic acceleration scenarios are respectively shown in Table \ref{tab:MAE case study I} and Fig. \ref{fig:MAE}. Each MAE is computed over the last 45 s of simulation and averaged over 1000 MC simulations. Table \ref{tab:MAE case study I} includes the MAE of the diagonal elements of the estimated $\mathbf{Q}$ except for ADMC since it is not directly comparable to the truth. While $\widetilde{Q}$ was initialized as one for the data in Table \ref{tab:MAE case study I}, Fig. \ref{fig:MAE} shows the MAE as a function of the initial guess of the process noise power spectral density $\widetilde{Q}_0$. For CM, the IMM technique, and the method developed by Moghe et al.\cite{moghe_adaptive_2019}, the initial $\mathbf{Q}$ provided to the filter is computed through Eq. (\ref{eq:Q SNC Case I}) using the specified $\widetilde{Q}_0$. The MAE when process noise is not modeled in the filter is also shown in Fig. \ref{fig:MAE} for reference. Not modeling process noise leads to filter inconsistency and large estimation error. SNC and DMC only perform well when a near optimal value of $\widetilde{Q}$ is used. In both scenarios, the developed algorithms significantly outperform CM and the technique developed by Moghe et al. in estimating position and its corresponding component of the process noise covariance $Q_{1,1}$. When using the method proposed by Moghe et al.\cite{moghe_adaptive_2019}, each element of the estimated $\mathbf{Q}$ converges to its true value in the stochastic acceleration scenario. However, in both scenarios most new estimates of $\mathbf{Q}$ provided by this technique are not positive definite, which leads to infrequent updates of the $\mathbf{Q}$ utilized in the filter. The IMM technique has the highest computational cost and its accuracy is dependent on how tightly the bounds on $\widetilde{Q}$ enclose the optimal value. As expected, ADMC only provides an advantage over ASNC when the unmodeled accelerations are correlated in time. Acceleration tracking and filter convergence plots for the deterministic acceleration scenario are provided in Figs. \ref{fig:1D filter convergence} and \ref{fig:1D acceleration tracking} in Appendix A. The average runtime per filter call when using a fixed $\mathbf{Q}$ was \SI{46}{\micro\second} for a MATLAB implementation on a 4 GHz Intel Core i7-6700 processor. The percent increase in filter computation time incurred by each adaptive filtering algorithm relative to a filter with a fixed $\mathbf{Q}$ is shown in Table \ref{tab:MAE case study I}.

\begin{table*}[!h]
\centering
\begin{threeparttable}
\caption{Case study I MAE for the stochastic acceleration scenario and percent increase in runtime.}
\begin{tabular}{l c c c c c c c}
\toprule
\toprule
Parameter													&Ideal$^*$ &CM							&IMM$^\dagger$ &IMM$^{\dagger\dagger}$ &Moghe et al.		&ASNC								&ADMC\\
\toprule
$x$ MAE (m) 											&0.121 								&0.505 							&0.122 						 &0.186							&0.475								&0.123 							&0.122\\
$\dot{x}$ MAE (m/s)								&7.38$\times 10^{-2}$		&7.43$\times 10^{-2}$  	&7.58$\times 10^{-2}$ &1.08$\times 10^{-1}$	&7.40$\times 10^{-1}$	&7.42$\times 10^{-2}$ 	&7.70$\times 10^{-2}$\\
$Q_{1,1}$ MAE (m)\textsuperscript{2}		&0										&3.09$\times 10^{-2}$  	&1.28$\times 10^{-4}$ &9.57$\times 10^{-3}$	&1.56$\times 10^{-1}$	&4.46$\times 10^{-5}$	&-\\
$Q_{2,2}$ MAE (m/s)\textsuperscript{2}	&0										&1.18$\times 10^{-2}$   &3.84$\times 10^{-2}$ &2.87 							&4.45$\times 10^{-3}$	&1.34$\times 10^{-2}$	&-\\		
Runtime ($\%$)	&0	&25	&223 &223	&41	 &74	 &79\\
\bottomrule
\bottomrule
\end{tabular}
\label{tab:MAE case study I}
\begin{tablenotes}
      \footnotesize
      \item \hspace{-.1cm}$^*$Optimal non-adaptive Kalman filter where the modeled $\mathbf{Q}$ matches the reference truth.
      \item \hspace{-.1cm}$^\dagger$ $\widetilde{Q}_{min} = 10^{-2}$ m$^2$/s$^{3}$ and $\widetilde{Q}_{max} = 1$ m$^2$/s$^{3}$.
      \item \hspace{-.1cm}$^{\dagger \dagger}$ $\widetilde{Q}_{min} = 10^{-3}$ m$^2$/s$^{3}$ and $\widetilde{Q}_{max} = 100$ m$^2$/s$^{3}$.
\end{tablenotes}
\end{threeparttable}
\end{table*}

\section{Case Study II - Formation Flying About an Asteroid}
This section further validates the developed process noise techniques by applying them to an idealized simulation of the autonomous navigation of two spacecraft orbiting the asteroid 433 Eros, which was characterized by the NEAR mission \cite{miller_determination_2002,konopliv_global_2002,williams_technical_2002}. More recent asteroid missions \cite{lauretta_osirisrex_2017,hashimoto_vision-based_2010} demonstrate a continued interest in asteroids due to several reasons including scientific value, mining, and planetary defense\cite{council_vision_2011,coradini_vesta_2011,takahashi_gravity_2013,mazanek_asteroid_2015}. Due to light time delay and limited ground-based resources such as NASA's Deep Space Network, there have been efforts to increase spacecraft autonomy for asteroid missions\cite{leonard_absolute_2012,hesar_small_2015,bhaskaran_small_2011,fujimoto_stereoscopic_2016,stacey_autonomous_2018}. The proposed process noise techniques enhance spacecraft autonomy and are well suited for the challenges of asteroid missions such as limited a priori knowledge of the dynamical environment and time-varying process noise statistics. 

For this case study, the chief and deputy spacecraft are both in near-circular, slightly retrograde orbits with an initial osculating semi-major axis of 40 km. In order to achieve passive collision avoidance between spacecraft, E/I vector separation is used to select the initial osculating relative orbital elements of the deputy\cite{damico_autonomous_2010}. The reference truth dynamics include Eros gravity up to degree and order 15 \cite{konopliv_global_2002} , sun third body effects, and solar radiation pressure assuming a constant area to mass ratio and reflectivity coefficient of each spacecraft. Every five minutes, interspacecraft radio-frequency range and range-rate measurements as well as spacecraft camera pixel measurements of optical navigation (OpNav)\cite{hesar_small_2015,gaskell_characterizing_2008,bhaskaran_small_2011} features on the asteroid surface are generated. The measurements are corrupted by zero-mean Gaussian white noise with standard deviations of 10 cm, 1 mm/s, and 0.5 pixels respectively. The considered measurement types are consistent with the recently proposed ANS mission architecture\cite{ANS,stacey_autonomous_2018}. Further details of the reference truth orbit geometry, dynamics, and measurement models are provided in Appendix B. 

A UKF\cite{thrun_probabilistic_2005} estimates the Cartesian states of the chief and deputy spacecraft. The technique developed by Moghe et al.\cite{moghe_adaptive_2019} cannot be applied because it is limited to linear time-invariant systems. The performance of ASNC, ADMC, IMM, and the CM technique given by Eq. (\ref{eq:Qhat estimate ss}) are each characterized through 1000 MC simulations in each of three scenarios. The three scenarios will be referred to as the no-maneuver, perfect-maneuver, and imperfect-maneuver scenarios. In the no-maneuver scenario, the spacecraft are only subject to the natural dynamics of the system, and the process noise is due to a truncated filter dynamics model. The only difference in the perfect-maneuver scenario is that the chief spacecraft performs a maneuver after 3.2 orbit periods that is modeled perfectly in the filter. This scenario serves as a baseline for the imperfect-maneuver scenario where the nominal maneuver modeled by the filter does not match the reference truth due to thruster uncertainties. The imperfect maneuver briefly creates unmodeled accelerations that are often several times larger than those due to the natural dynamics. In each MC simulation, the initial estimate error in position and velocity is randomly sampled from a zero-mean normal distribution whose covariance is the initial formal covariance provided to the filter. Each simulation lasts for four orbit periods where an orbit period is approximately 20.9 hours. 

\subsection{Filter}
The UKF estimated state when using CM, IMM, or ASNC is
\begin{equation} \label{eq:state}
\bm{x} = [\bm{r}_c^T \ \ \bm{v}_c^T \ \ \bm{r}_d^T \ \ \bm{v}_d^T]^T
\end{equation}
where $\bm{r}$ and $\bm{v}$ indicate spacecraft position and velocity vectors respectively expressed in the Asteroid Centered Inertial (ACI) frame, which is aligned with the J2000 reference frame. The subscripts $c$ and $d$ respectively denote the chief and deputy spacecraft. When using ADMC, the state is augmented with a set of three empirical accelerations expressed in the ACI frame for each spacecraft, and the state vector becomes
\begin{equation} \label{eq:state ADMC}
\bm{x} = [\bm{r}_c^T \ \ \bm{v}_c^T \ \ \bm{\tilde{a}}_c^T \ \ \bm{r}_d^T \ \ \bm{v}_d^T \ \ \bm{\tilde{a}}_d^T]^T
\end{equation}
In the time update, the filter propagates the state using a fourth-order Runge-Kutta numerical integration of the modeled spacecraft dynamics. In the no-maneuver scenario, the filter model of the spacecraft dynamics only takes into account the two-body gravity and $J_2$ of the asteroid, which are assumed to have been determined earlier in the mission. In the perfect-maneuver scenario, the filter dynamics model also includes a nominal chief maneuver that matches the reference truth. In the imperfect-maneuver scenario, the errors in the filter modeled maneuver magnitude and direction are randomly sampled in each MC simulation. The error in maneuver magnitude is sampled from a zero-mean normal distribution with a standard deviation equal to 15$\%$ of the true maneuver magnitude. The error in maneuver direction is simulated by rotating the unit vector of the true maneuver direction through a 3-2-1 Euler Angle rotation where each of the three rotation angles are independently sampled from a zero-mean normal distribution with a standard deviation of 0.5$\degree$. The filter measurement models are consistent with the reference truth as shown in Eqs. (\ref{eq:RF meas}) and (\ref{eq:pixel meas2}) in Appendix B. The rotation matrices used to compute pixel measurements in Eq. (\ref{eq:pixel meas2})  are assumed to be known from characterization of the asteroid rotational motion earlier in the mission and an onboard star tracker. In each simulation, the initial 1-$\sigma$ formal uncertainties provided to the filter in position and velocity in each axis are 1 km and 50 mm/s. The exploiting triangular structure (ETS) technique previously developed by the authors is used in the UKF time update to reduce computation time with no loss of accuracy by reusing spacecraft orbit propagations when propagating sigma points\cite{stacey_autonomous_2018}.

CM, ASNC, and ADMC are each utilized with a sliding window of 30 time steps, which is a 2.5 hour interval. Since the filter initially converges very rapidly, adaptation of the process noise covariance is delayed by ten measurement intervals to avoid including data in the sliding window average that significantly violates the steady state assumption in CM, ASNC, and ADMC. This improves initial filter convergence behavior, but does not change steady state performance. The weighting matrix for ASNC and ADMC is approximated as diagonal using Eq. (\ref{eq:weighting matrix}), which allows for the computationally efficient estimate of $\mathbf{\widetilde{Q}}$ shown in Eq. (\ref{eq:sol ASNC}). No upper bound on $\mathbf{\widetilde{Q}}^{diag}$ is utilized for the developed algorithms, and the lower bound is set to a vector of zeros. For ASNC, the model of the process noise covariance is approximated using Eq. (\ref{eq:Q ASNC efficient}). For ADMC, the approximate model of $\mathbf{Q}$ in Eq. (\ref{eq:Q ADMC efficient}) is used, the empirical accelerations are initialized as zero, $\alpha$ is set to 0.02, and $\bm{\beta}$ is a diagonal matrix with each element on the diagonal equal to 1$\times 10^{-5}$ s\textsuperscript{-1}. For IMM, the modeled $\mathbf{Q}$ is computed through Eq. (\ref{eq:Q ASNC efficient}) where the lower and upper bounds on $\mathbf{Q}$ model each diagonal element of $\mathbf{\widetilde{Q}}$ as $10^{-8}$ m$^2$/s$^{3}$ and $10^{-6}$ m$^2$/s$^{3}$ respectively. These bounds contain the optimal value, which was determined to be approximately $10^{-7}$ m$^2$/s$^{3}$ through manual tuning. The initial probability of each mode is set as 0.5 and the transition probabilities are defined according to Eq. (\ref{eq:transition probabilities}).

\subsection{Results}
\begin{figure*}[!h]
\centering 
\subfigure[CM]{\label{fig:case study II filter convergence a}
\includegraphics[width=.22\linewidth,trim= 0 0 0 0,clip]{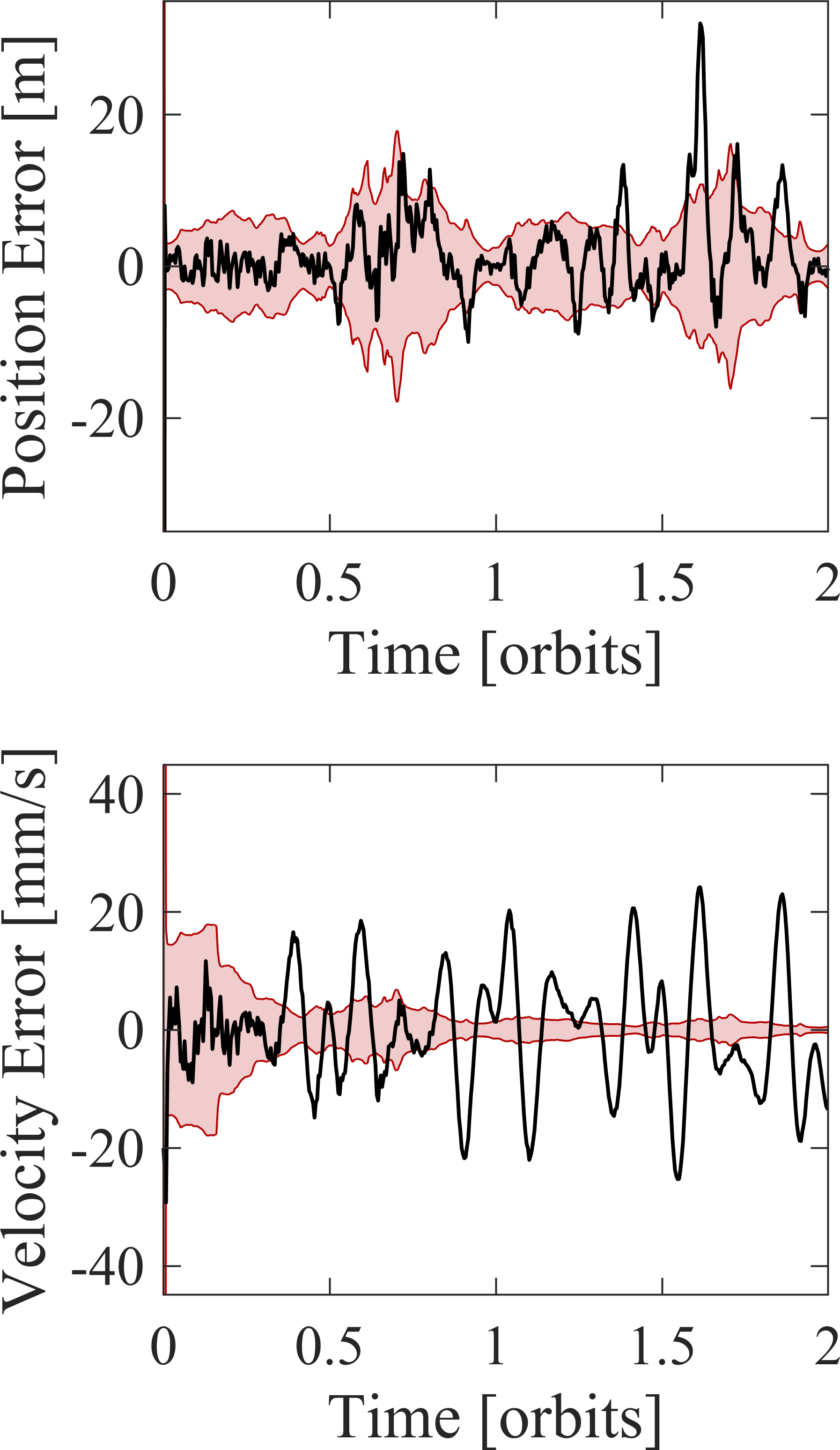}}
\subfigure[IMM]{\label{fig:case study II filter convergence d}
\includegraphics[width=.22\linewidth,trim= 0 0 0 0,clip]{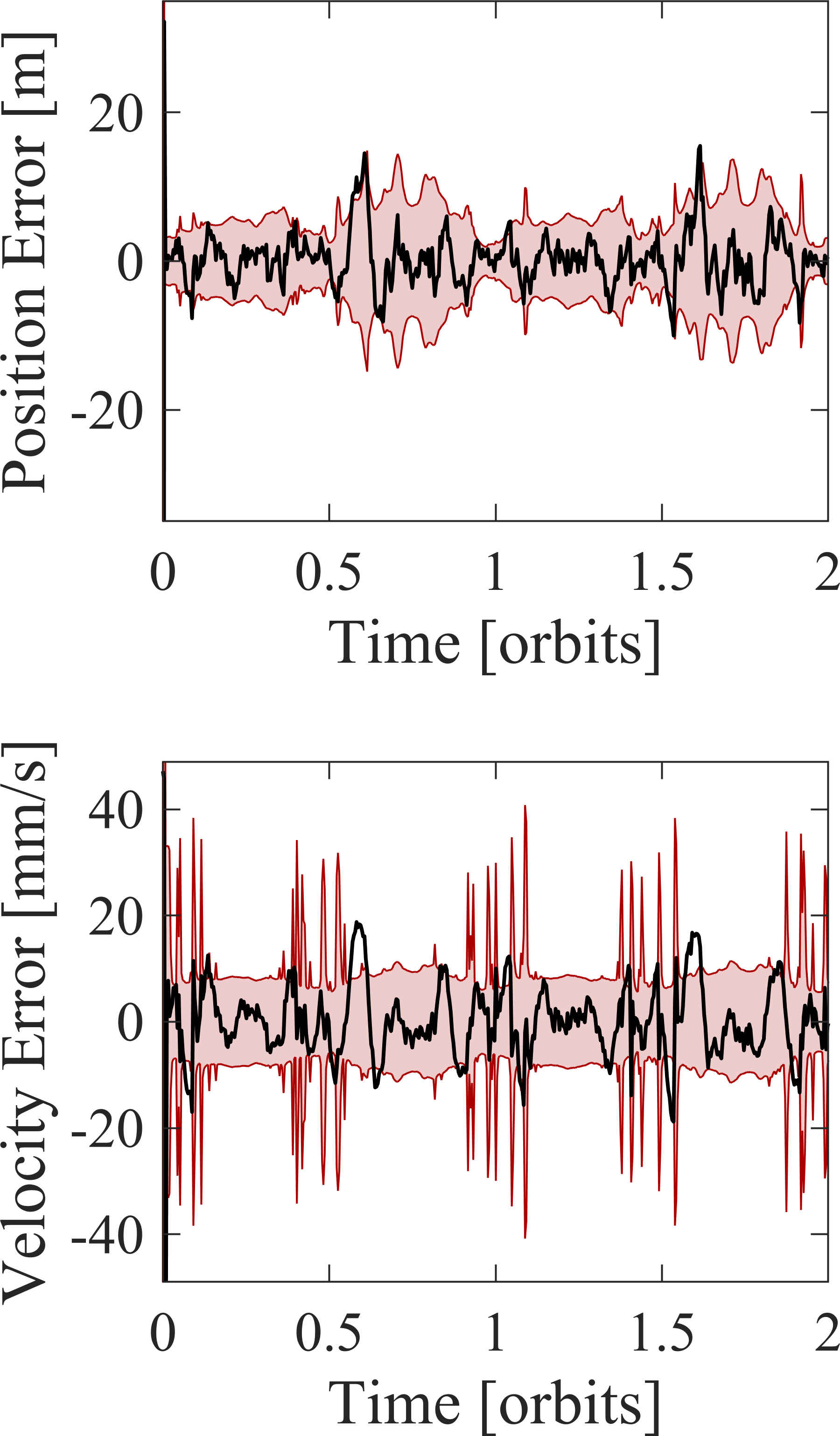}}
\subfigure[ASNC]{\label{fig:case study II filter convergence b}
\includegraphics[width=.22\linewidth,trim= 0 0 0 0,clip]{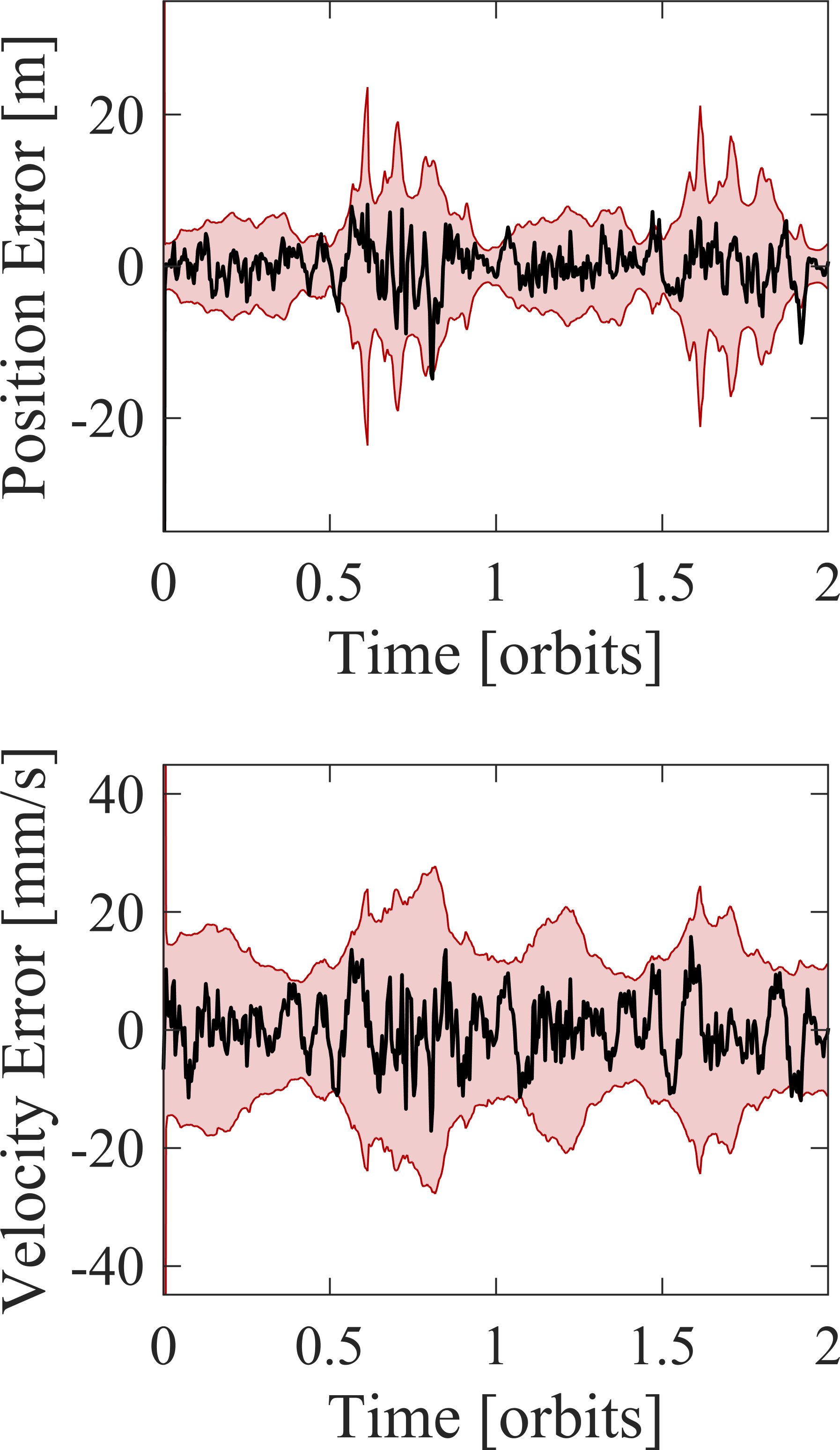}}
\subfigure[ADMC]{\label{fig:case study II filter convergence c}
\includegraphics[width=.22\linewidth,trim= 0 0 0 0,clip]{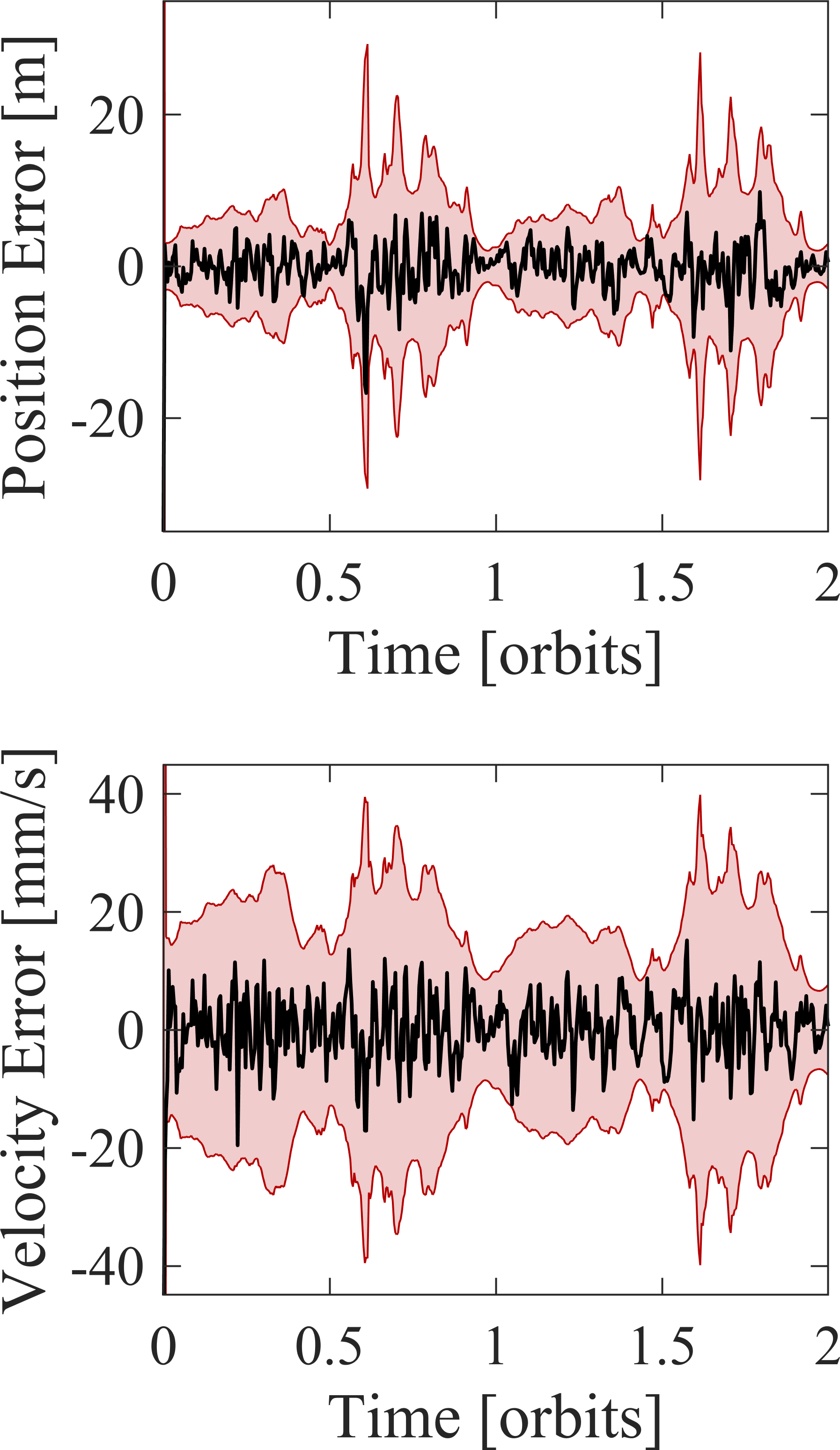}}
\caption{Estimation error (black) and corresponding formal 3-$\sigma$ bound (red) of the x-component of the chief position (top row) and velocity (bottom row) vectors for case study II.}
\label{fig:case study II filter convergence}
\end{figure*}

Initial filter convergence is shown in Fig. \ref{fig:case study II filter convergence} for the x-component of the chief position and velocity vectors. Filter convergence is similar in all three axes and for each spacecraft. Fluctuations in the formal uncertainty are largely due to the time-varying number of visible OpNav landmarks. As can be seen in Fig. \ref{fig:case study II filter convergence a}, the inherently biased estimate of $\mathbf{Q}$ provided by CM causes filter inconsistency. Maneuver error further degrades the performance of CM as shown in Fig. \ref{fig:case study II imperfect maneuver filter convergence} in Appendix C. In contrast IMM, ASNC, and ADMC provide consistent estimation.

The mean 3D error computed over the last two orbits and averaged over 1000 MC simulations for each scenario is listed in Table \ref{tab:MAE} for each of the considered process noise techniques. For the no-maneuver scenario, ASNC provided 47\% less error in position and 51\% less error in velocity than CM. ADMC reduced position error by 19\% and velocity error by 9\% over ASNC. Navigation performance for IMM, ASNC, and ADMC improved in the perfect-maneuver scenario due to the change in orbit geometry. Introducing error in the maneuver only slightly decreased navigation accuracy for IMM, ASNC, and ADMC but significantly degraded performance for CM. In fact, 12 of the MC simulations for CM in the imperfect-maneuver scenario diverged. The performance of IMM depends on the bounds on $\mathbf{Q}$ but provides slightly worse performance than ASNC for the chosen bounds. Unmodeled acceleration tracking for ADMC is shown in Fig. \ref{fig:case study II acceleration tracking} in Appendix C. Fig. \ref{fig:xacelleration tracking imperfect maneuver} demonstrates that ADMC is often able to continue tracking the unmodeled accelerations during an imperfect maneuver, even though the unmodeled accelerations are suddenly much larger for only three measurement intervals. 

\begin{table}[!h]
\centering
\begin{threeparttable}
\caption{Case study II mean 3D errors.}
\begin{tabularx}{260pt}{l l *{4}{>{\Centering}X}}
\toprule
\toprule
Test Scenario 			&Error Type				&CM		&IMM 	&ASNC	&ADMC\\
\toprule
No-maneuver 			&Position (m) 			&9.14  	&4.98			&4.82 	&3.89\\
 								&Velocity (mm/s)		&19.1 	&10.0			&9.41	&8.59\\
\midrule
Perfect-maneuver 	&Position (m) 			&12.5  	&4.74			&4.55	&3.78\\
 								&Velocity (mm/s)		&18.8	&9.63			&8.84	&8.26\\
\midrule
Imperfect-maneuver &Position (m) 			&21.4$^*$ &4.75			&4.58	&3.80\\
 								 &Velocity (mm/s)		&109$^*$	&9.68			&8.96	&8.49\\ 		
\bottomrule
\bottomrule
\end{tabularx}
\label{tab:MAE}
\begin{tablenotes}
      \footnotesize
      \item \hspace{-.1cm}$^*$Does not include the 12 MC simulations where the filter diverged.
\end{tablenotes}
\end{threeparttable}
\end{table}

Filter computation time is dominated by propagating the $2n+1$ sigma points in the time update where $n$ is the number of state variables. For the states in Eqs. (\ref{eq:state}) and (\ref{eq:state ADMC}), there are 25 and 37 sigma points respectively. Since the estimated state comprises two spacecraft states, a traditional UKF would propagate $2(2n+1)$ spacecraft orbits at each filter call. However, the ETS technique reduces the number of orbit propagations from 50 to 38 for the state in Eq. (\ref{eq:state}) and from 74 to 56 for the augmented state in Eq. (\ref{eq:state ADMC}). As a result, ETS reduces filter runtime by approximately 24\% regardless of which process noise technique is used. The average runtime per filter call for the ETS-UKF when using a fixed $\mathbf{Q}$ was 66 ms for a MATLAB implementation on a 4 GHz Intel Core i7-6700 processor. CM and ASNC only incur an additional 3.4$\times10^{-3}$\% and 1.0$\times10^{-2}$\% in runtime respectively. ADMC increases computation time by 47\% due to the increased number of sigma point propagations, and the IMM method increases computation time by 100\% due to the second filter running in parallel.

\section{Conclusions}
This paper presents two new techniques to accurately estimate the process noise covariance of a discrete-time Kalman filter online for robust orbit determination in the presence of dynamics model uncertainties. Key limitations of state noise compensation (SNC), dynamic model compensation (DMC), and existing adaptive filtering approaches are overcome by optimally fusing SNC and DMC with covariance matching (CM) adaptive filtering. This yields two new techniques called adaptive SNC (ASNC) and adaptive DMC (ADMC). The adaptability of the developed algorithms is a significant advantage over SNC and DMC, which require onerous offline tuning. Unlike SNC and DMC, the new techniques are applicable when the process noise statistics are time-varying and the dynamical environment is not well known a priori, which is typical for asteroid missions. In contrast to many adaptive filtering approaches, ASNC and ADMC are suitable for onboard orbit determination because they are computationally efficient, do not assume a linear time-invariant system, and update the filter process noise covariance online. Additionally, the new algorithms guarantee the process noise covariance is positive semi-definite  without relying on ad hoc methods and constrain the process noise covariance according to the underlying continuous-time dynamical model. The developed algorithms also accurately extrapolate the process noise covariance over measurement outages and can easily leverage a priori bounds on the process noise power spectral density. Furthermore, ADMC is well suited for systems with colored process noise. The advantages of the developed algorithms are first demonstrated through an illustrative linear system and then through a nonlinear system comprising the idealized navigation of two spacecraft orbiting an asteroid. As shown in these case studies, ASNC is less computationally expensive and simpler to implement than ADMC. ASNC also has a single tunable parameter, the length of the sliding window, while in the present work ADMC additionally requires selection of a forgetting factor and the empirical acceleration time correlation constants. On the other hand, ADMC can provide superior estimation when the measurement rate and accuracy are sufficient to enable accurate tracking of the unmodeled accelerations. The proposed algorithms have the potential to improve orbit determination for a variety of missions and are especially relevant for missions to small celestial bodies such as asteroids.

\newpage
\appendices
\onecolumn
\section{Case Study I Additional Filter Performance Plots}
\begin{figure*}[!h]
\centering 
\subfigure[No modeled $\mathbf{Q}$ and CM]{\label{fig:1D filter convergence a}
\includegraphics[width=.29\linewidth,trim= -10 0 -10 0,clip]{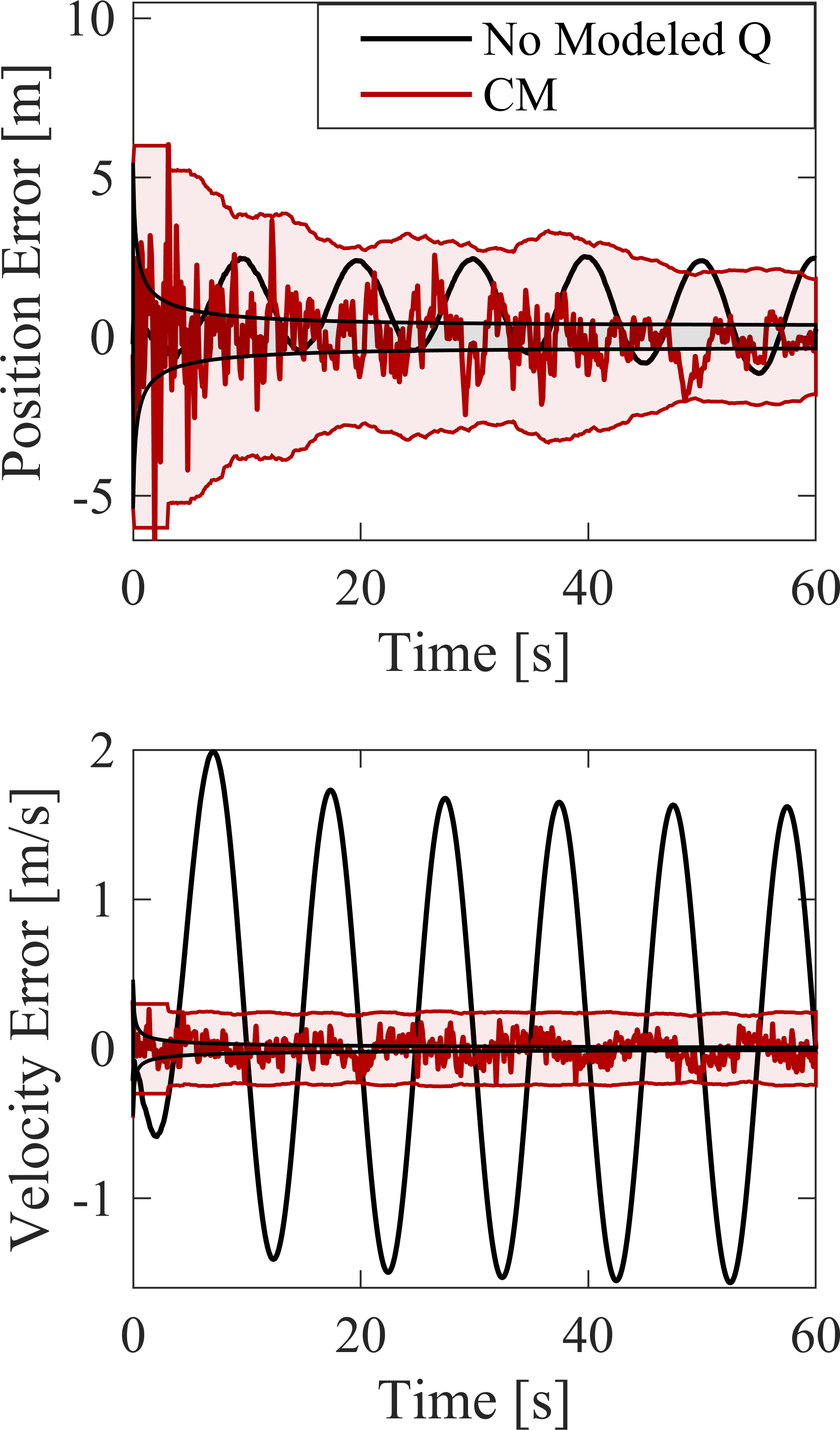}}
\subfigure[SNC and ASNC]{\label{fig:1D filter convergence b}
\includegraphics[width=.29\linewidth,trim= -10 0 -10 0,clip]{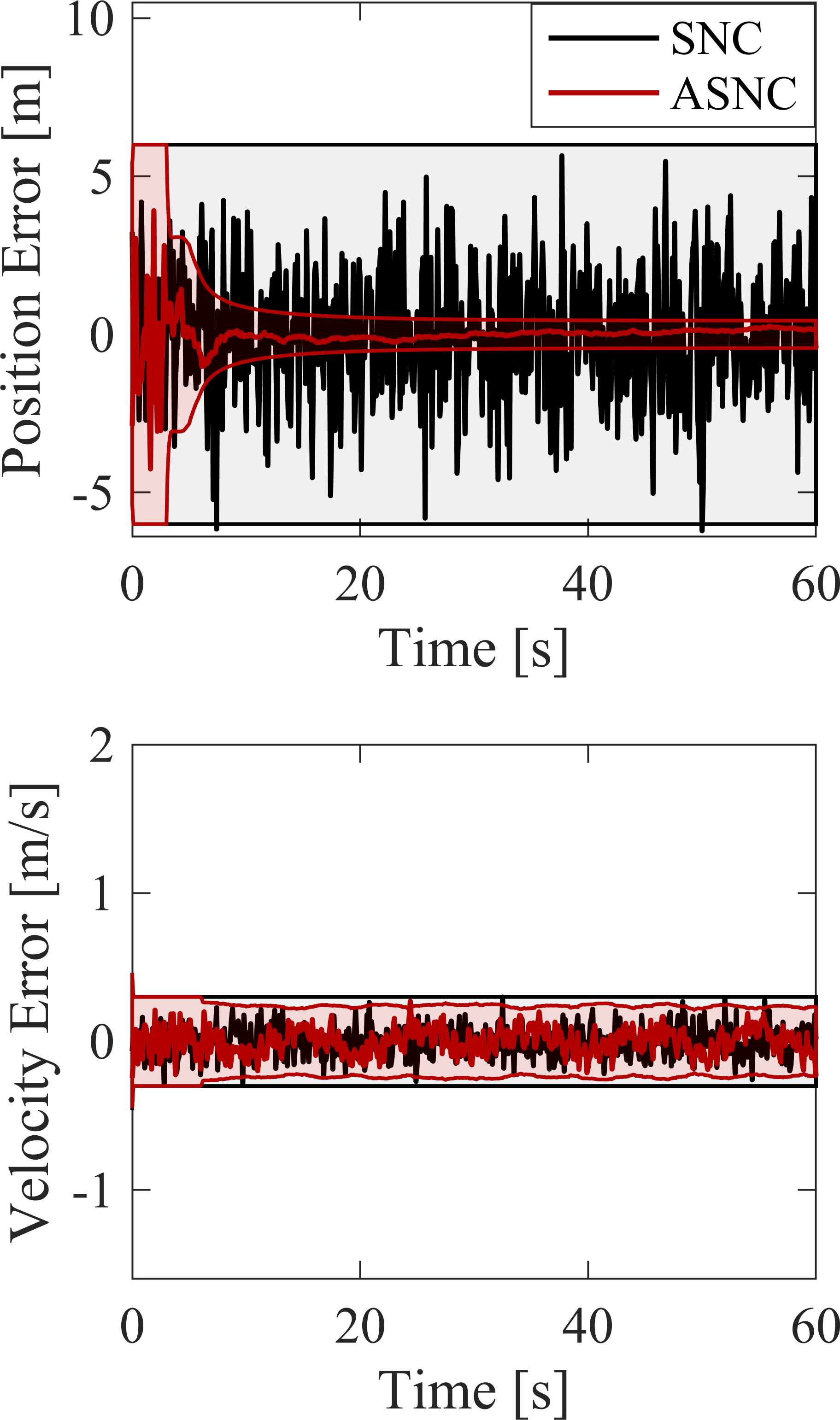}}
\subfigure[DMC and ADMC]{\label{fig:1D filter convergence c}
\includegraphics[width=.29\linewidth,trim= -10 0 -10 0,clip]{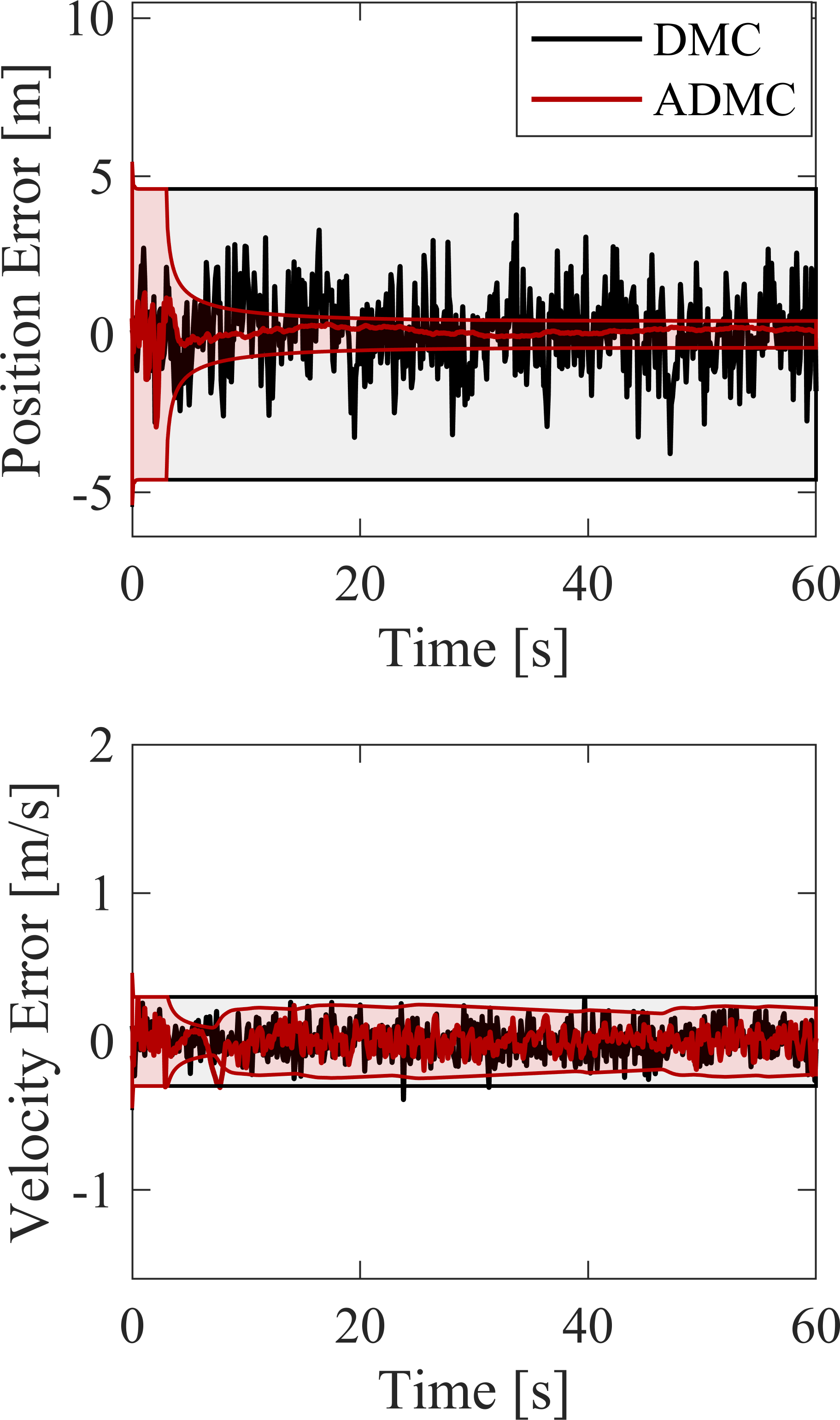}}
\caption{True error for the deterministic acceleration scenario where the shaded region of the same color is the corresponding formal 3-$\sigma$ bound. Each process noise technique is initialized with a much larger than optimal $\widetilde{Q}_0 = 10^8$ in the corresponding units of either m$^2$/s$^{3}$ or m$^2$/s$^{5}$.}
\label{fig:1D filter convergence}
\end{figure*}

\begin{figure*}[!h]
\centering 
\subfigure[$\widetilde{Q}_0 = 10^{-12}$ m$^2$/s$^{5}$]{\label{fig:1D acceleration tracking a}
\includegraphics[width=.29\linewidth,trim= -5 0 -5 0,clip]{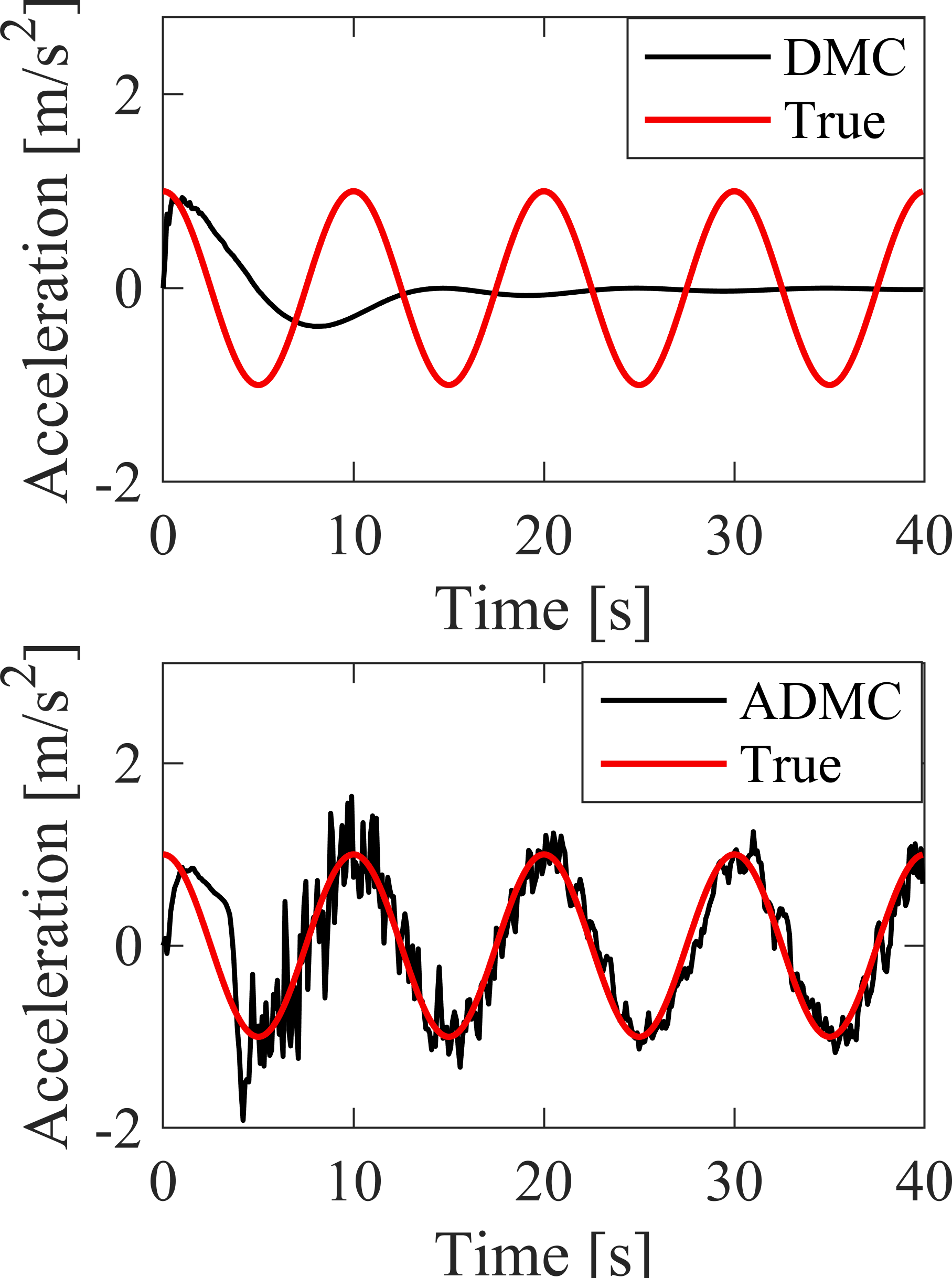}}
\subfigure[$\widetilde{Q}_0 =0.206$ m$^2$/s$^{5}$]{\label{fig:1D acceleration tracking b}
\includegraphics[width=.29\linewidth,trim= -5 0 -5 0,clip]{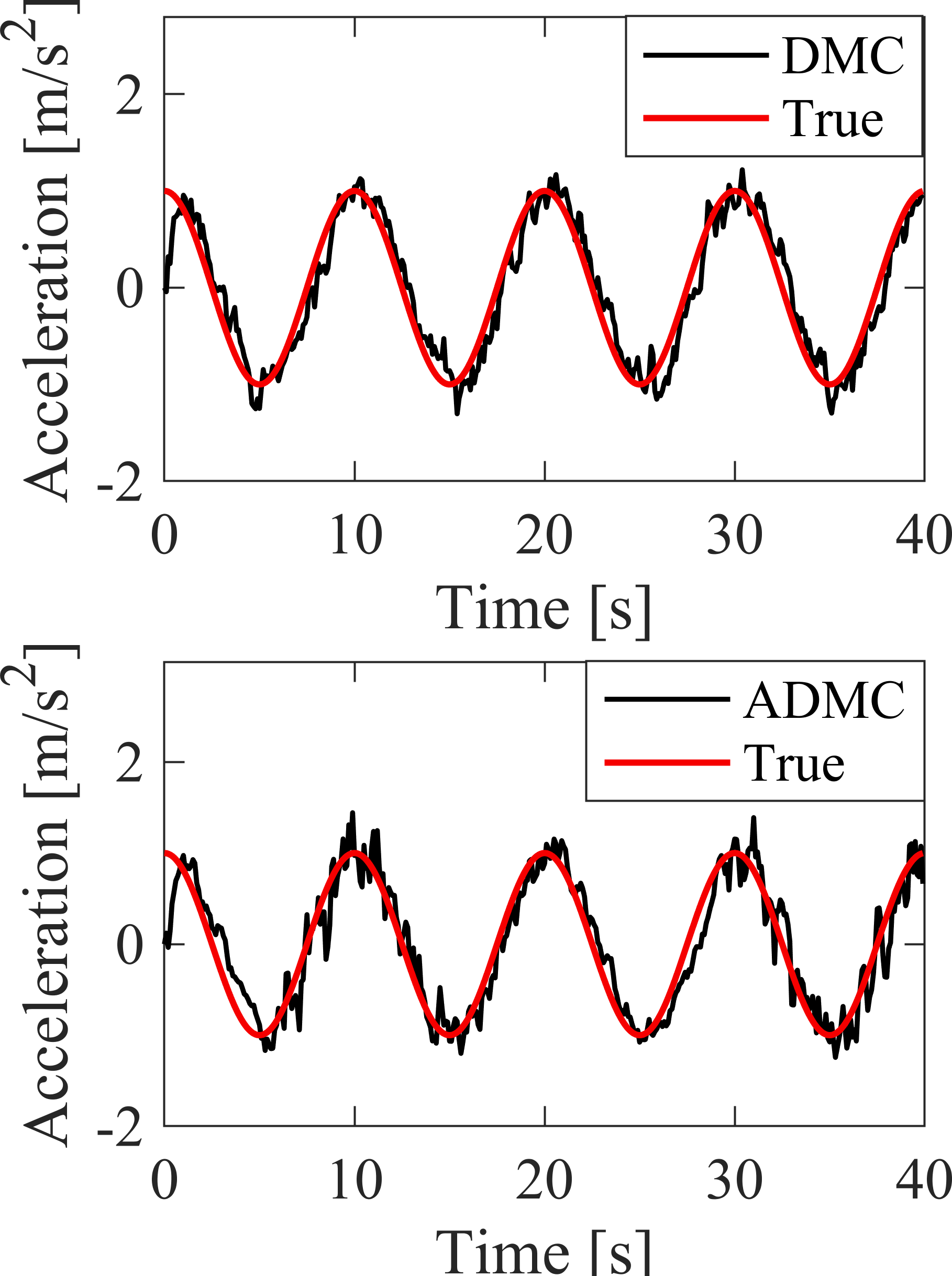}}
\subfigure[$\widetilde{Q}_0 = 10^8$ m$^2$/s$^{5}$]{\label{fig:1D acceleration tracking c}
\includegraphics[width=.29\linewidth,trim= -5 0 -5 0,clip]{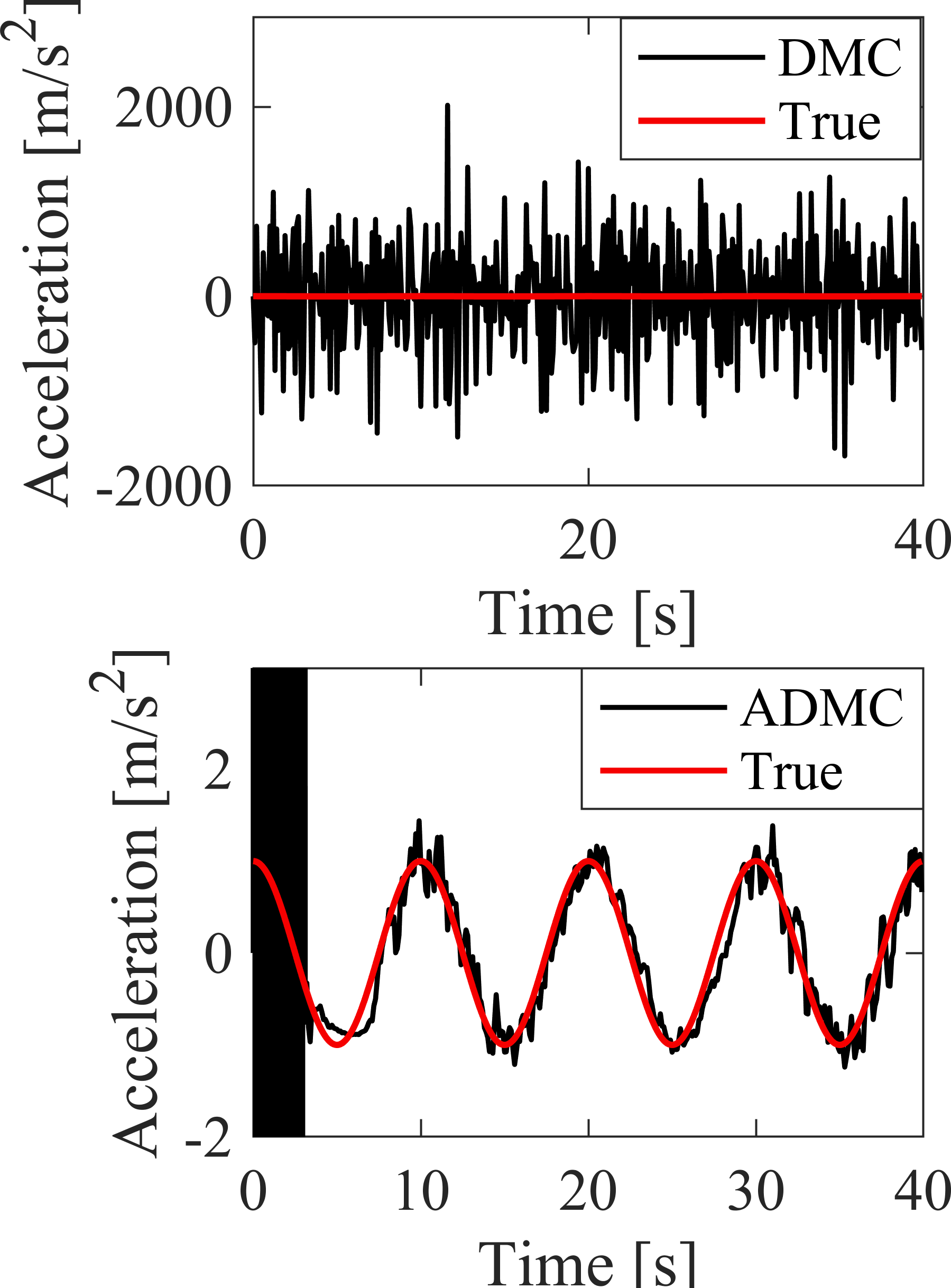}}
\caption{Comparison of the true deterministic unmodeled acceleration and estimated empirical acceleration for DMC (top row) and ADMC (bottom row). For DMC, $\widetilde{Q}_0 =0.206$ m$^2$/s$^{5}$ is optimal.}
\label{fig:1D acceleration tracking}
\end{figure*}

\twocolumn
\section{Case Study II Reference Truth Details}
The initial osculating Keplerian orbital elements of the chief are $[a_c\ e_c\ i_c\ \Omega_c\  \omega_c\ M_c] = [40\text{ km}\ 0.01\ 95\degree\ 0\degree\ 0\degree\ 0\degree]$. These orbital elements are defined with respect to an inertial frame centered at the asteroid center of mass. The z-axis of this frame is aligned with the mean spin axis of the asteroid, the x-axis is aligned with the asteroid prime meridian at the epoch J2000, and the y-axis completes the right-handed triad. In order to achieve passive collision avoidance between spacecraft, E/I vector separation is used to select the initial quasi-nonsingular relative orbital elements (ROE) of the deputy\cite{damico_autonomous_2010}. The ROE are defined in terms of the chief and deputy Keplerian orbital elements as\cite{damico_autonomous_2010,sullivan_nonlinear_2017,koenig_new_2017}
\begin{equation} \label{eq:roe}
\begin{bmatrix}
\delta a\\
\delta \lambda\\
\delta e_x\\
\delta e_y\\
\delta i_x\\
\delta i_y
\end{bmatrix}
=
\begin{bmatrix}
(a_d-a_c)/a_c\\
u_d-u_c+(\Omega_d-\Omega_c)\text{cos}(i_c)\\
e_d\text{cos}(w_d)-e_c\text{cos}(w_c)\\
e_d\text{sin}(w_d)-e_c\text{sin}(w_c)\\
i_d-i_c\\
(\Omega_d-\Omega_c)\text{sin}(i_c)
\end{bmatrix}
\end{equation}
Here $u=M+w$ is the mean argument of latitude, and the subscripts $c$ and $d$ indicate the chief and deputy respectively. The initial osculating ROE of the deputy multiplied by the chief semi-major axis are $a_c[\delta a\ \delta\lambda\ \delta e_x\ \delta e_y\ \delta i_x\ \delta i_y] = [0\ 5\ 0\ 2\ 0\ 2]\text{ km}$.

In the perfect-maneuver and imperfect-maneuver scenarios, the chief executes a continuous thrust maneuver in the opposite direction of its angular momentum vector for 15 min. The maneuver begins at approximately 3.2 orbit periods when the chief reaches an argument of lattitude of 90$\degree$. The thrust magnitude is 5.76 mN, which is consistent with the Busek Micro Resistojet\cite{lemmer_propulsion_2017}. Assuming an 8 kg spacecraft, the thrust accelerates the chief at 720 \SI{}{\micro \meter}/s\textsuperscript{2}. The maneuver decreases $\Omega_c$ by approximately 11$\degree$ which increases $a_c\delta i_y$ by about 8 km.

Idealized interspacecraft radio-frequency range and range-rate measurements are simulated as
 \begin{equation}\label{eq:RF meas}
	\rho 			= ||\bm{\rho}||,\hspace{.5cm}
	\dot{\rho} = \frac{\bm{\dot{\rho}} \cdot \bm{\rho}}{||\bm{\rho}||}
\end{equation}
where $\bm{\rho} = \bm{r}_d-\bm{r}_c$ is the position vector of the deputy with respect to the chief. Idealized pixel measurements, $u$ and $v$, taken by cameras onboard the chief and deputy spacecraft of visible OpNav features on the asteroid surface are given by the pinhole camera model
\begin{equation}\label{eq:pixel meas2}
\begin{bmatrix}
uw\\
vw\\
w
\end{bmatrix}
= \hat{\mathbf K}\underset{ACI\rightarrow CF}{\mathbf R}\bigg[\underset{ACAF\rightarrow ACI}{\mathbf R} \ \ \ \ -\hspace{-.05cm}\bm r\bigg]
\begin{bmatrix}
\bm L\\
1
\end{bmatrix}
\end{equation} 
Here $\bm r$ is the spacecraft position with respect to the asteroid center of mass expressed in the ACI frame. The OpNav landmark position expressed in the Asteroid Centered Asteroid Fixed (ACAF) frame is denoted by $\bm L$. Vectors expressed in the ACAF frame are expressed in the ACI frame through multiplication with the rotation matrix \tiny$\raisebox{1.8pt}{$\underset{ACAF\rightarrow ACI}{}$}$  \hspace{-.87cm}\small $\raisebox{2pt}{$\mathbf R$}$ \hspace{.47cm} \normalsize \hspace{-.11cm}, which is computed from the asteroid rotational parameters \cite{konopliv_global_2002}. Vectors expressed in the ACI frame are expressed in the Camera Fixed (CF) frame through the rotation matrix \mbox{\tiny$\raisebox{1.8pt}{$\underset{ACI\rightarrow CF}{}$}$  \hspace{-.71cm}\small $\raisebox{2pt}{$\mathbf R$}$ \hspace{.47cm} \normalsize \hspace{-.2cm}}. The matrix of known camera intrinsic parameters $\hat{\mathbf K}$ is defined as
\begin{equation}
\hat{\mathbf K} =
\begin{bmatrix}
f_x	&0 	&c_x\\
0 	&f_y 	&c_y\\
0 	&0 	&1
\end{bmatrix}
\end{equation}
where $f_x$ and $f_y$ indicate the camera focal length divided by the pixel pitch in the CF frame x and y directions respectively. The vector $\bm c=[c_x \ c_y]^T$ denotes the principal point in units of pixels. The simulated camera properties are consistent with the OSIRIS-REx NavCam \cite{bos_touch_2018}. Each camera continuously points at the asteroid center of mass. A set of 100 points on the asteroid surface were selected as OpNav features. Pixel measurements of an OpNav feature taken from a spacecraft camera are only available if the feature is directly illuminated by the sun, is within the camera field of view, and is not blocked from the camera view by the asteroid surface. 

\clearpage
\newpage
\onecolumn
\section{Case Study II Additional Filter Performance Plots}
\begin{figure*}[!h]
\centering 
\subfigure[CM]{\label{fig:case study II imperfect maneuver filter convergence a}
\includegraphics[width=.235\linewidth,trim= 0 0 0 0,clip]{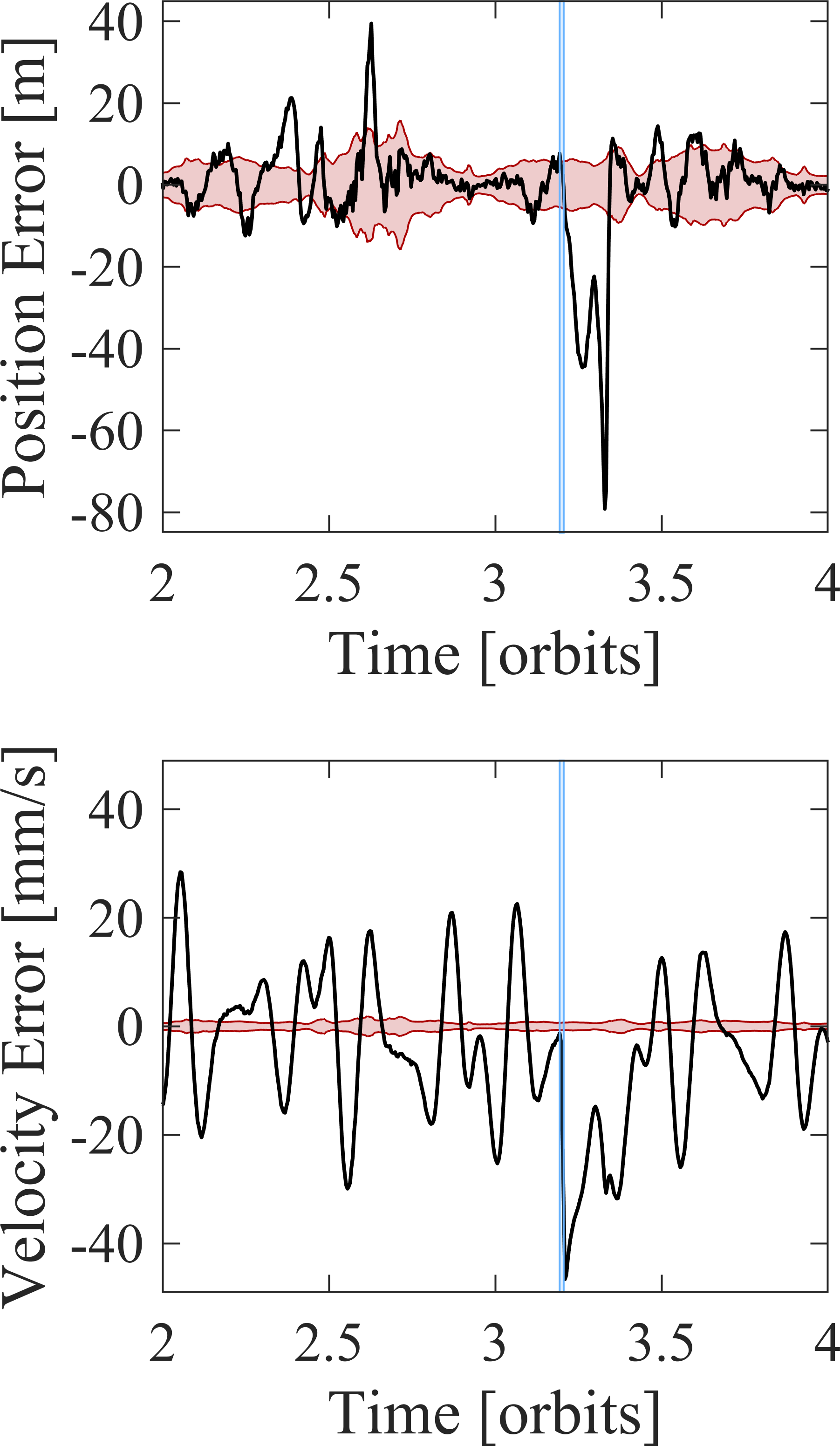}}
\subfigure[IMM]{\label{fig:case study II imperfect maneuver filter convergence d}
\includegraphics[width=.235\linewidth,trim= 0 0 0 0,clip]{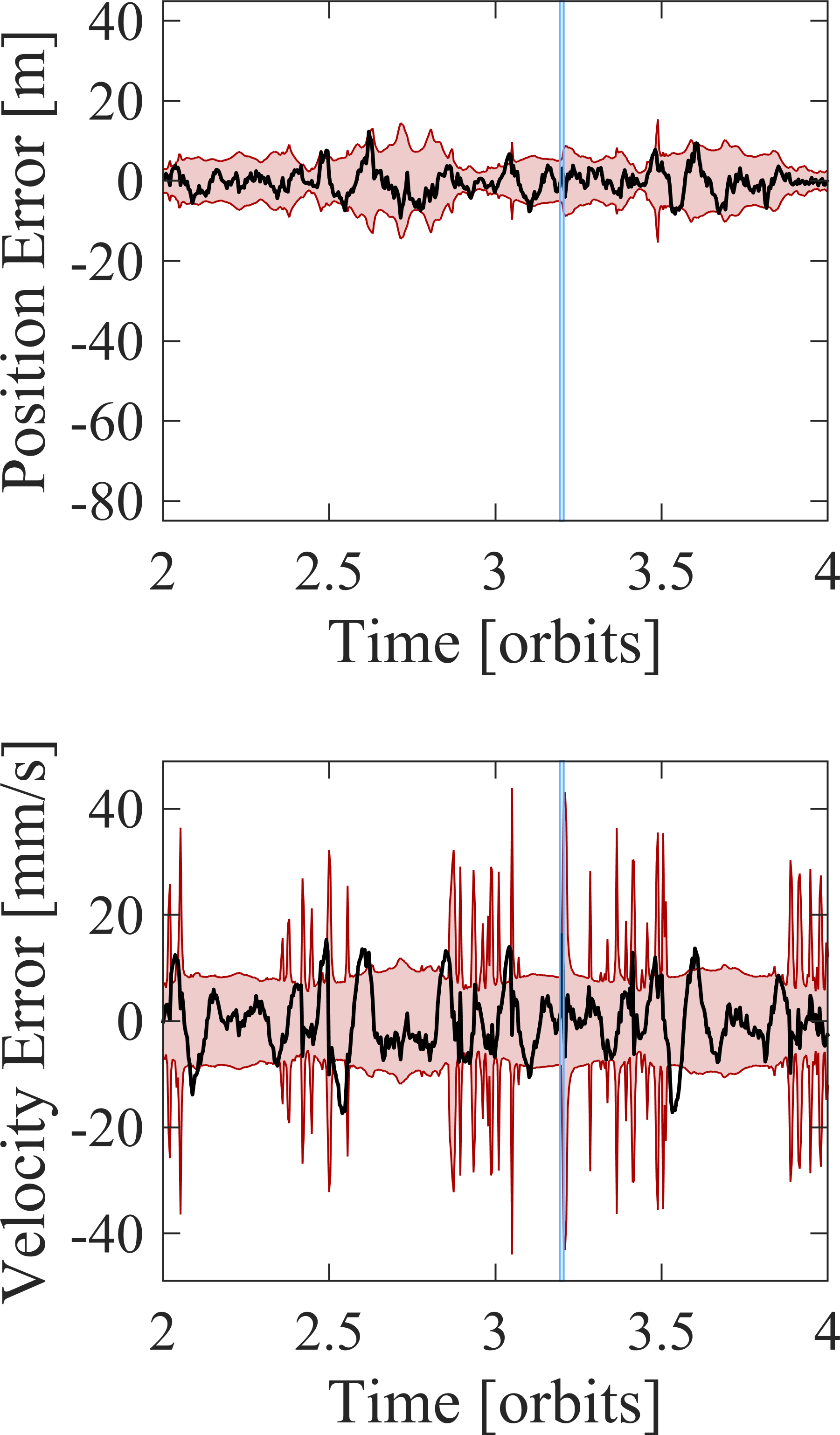}}
\subfigure[ASNC]{\label{fig:case study II imperfect maneuver filter convergence b}
\includegraphics[width=.235\linewidth,trim= 0 0 0 0,clip]{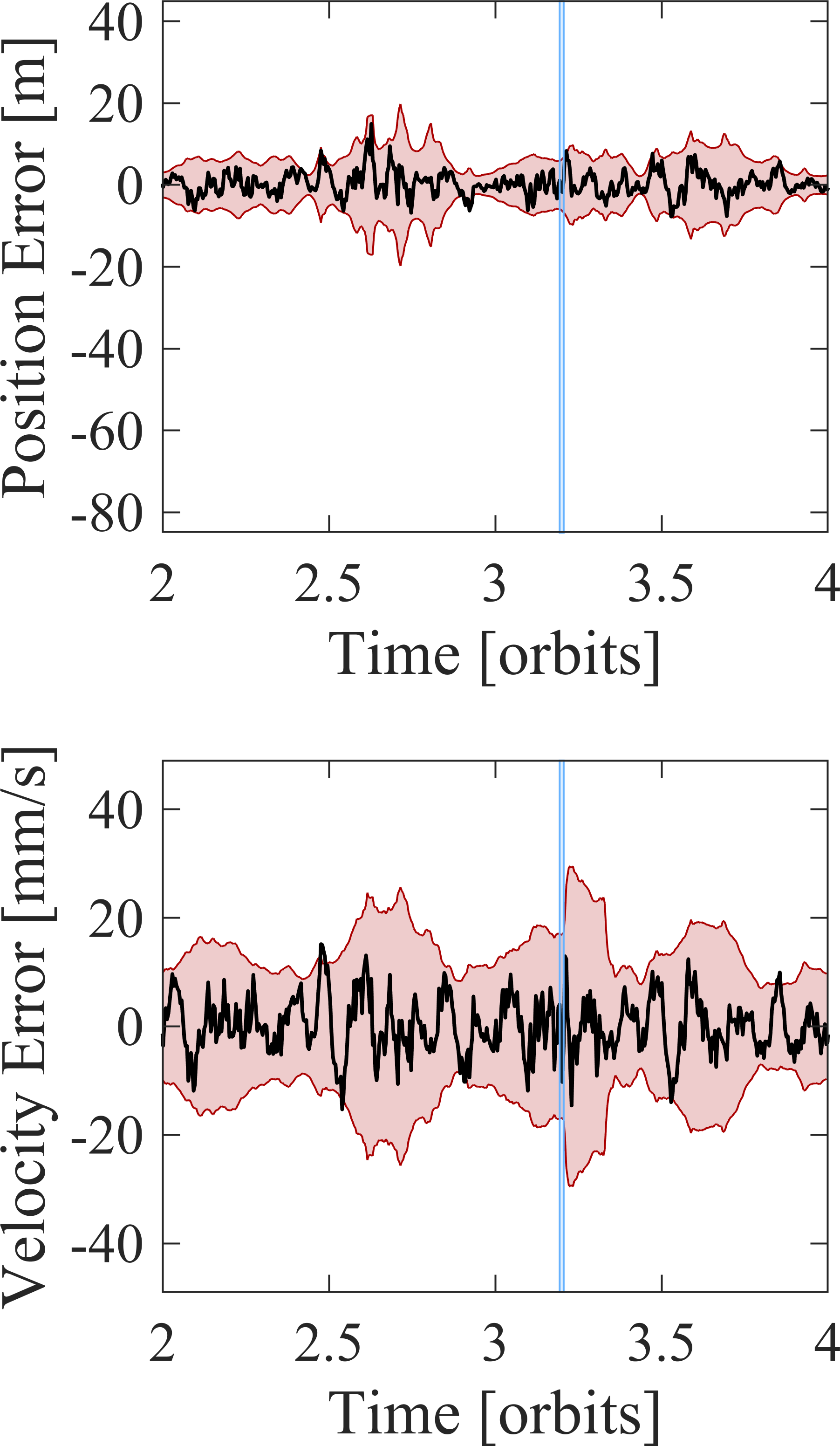}}
\subfigure[ADMC]{\label{fig:case study II imperfect maneuver filter convergence c}
\includegraphics[width=.235\linewidth,trim= 0 0 0 0,clip]{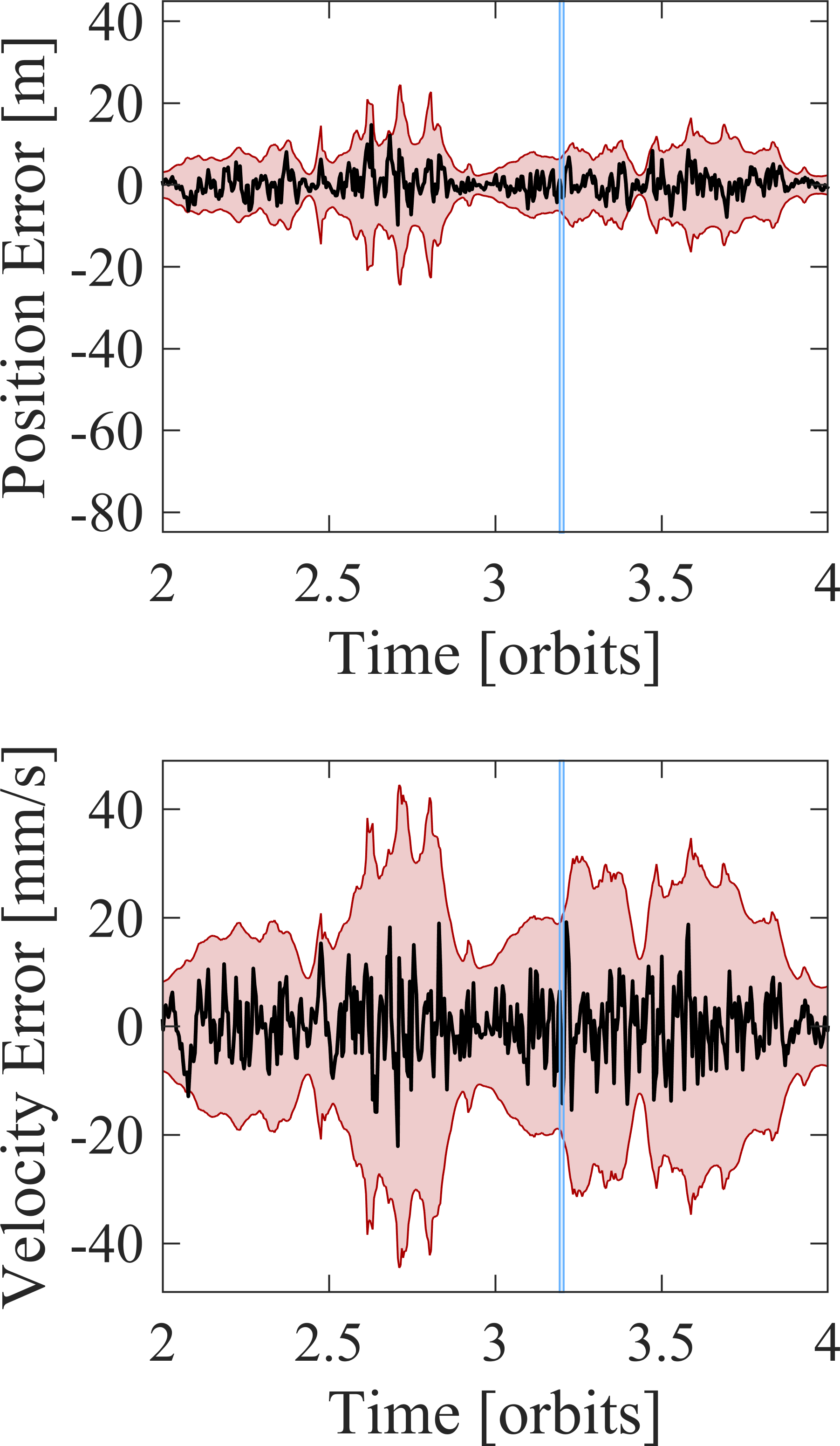}}
\caption{Imperfect-maneuver scenario estimation error (black) and corresponding formal 3-$\sigma$ bound (red) of the x-component of the chief position (top row) and velocity (bottom row) vectors where the blue region indicates the maneuver duration.}
\label{fig:case study II imperfect maneuver filter convergence}
\end{figure*}

\begin{figure*}[!h]
\centering 
\subfigure[No-maneuver]{\label{fig:xacelleration tracking}
\includegraphics[width=.48\linewidth,trim= 0 0 0 0,clip]{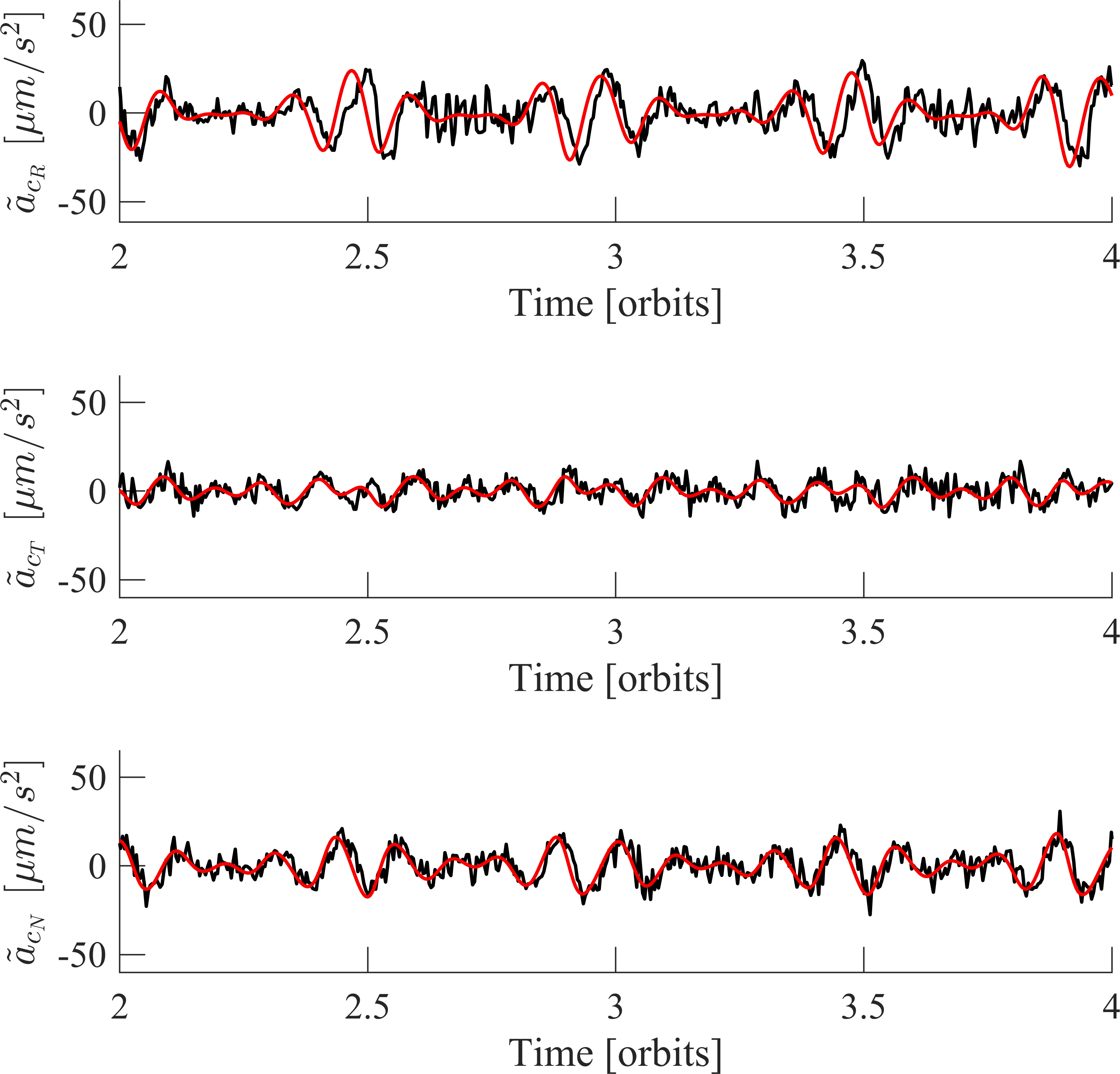}}
\subfigure[Imperfect-maneuver]{\label{fig:xacelleration tracking imperfect maneuver}
\includegraphics[width=.48\linewidth,trim= 0 0 0 0,clip]{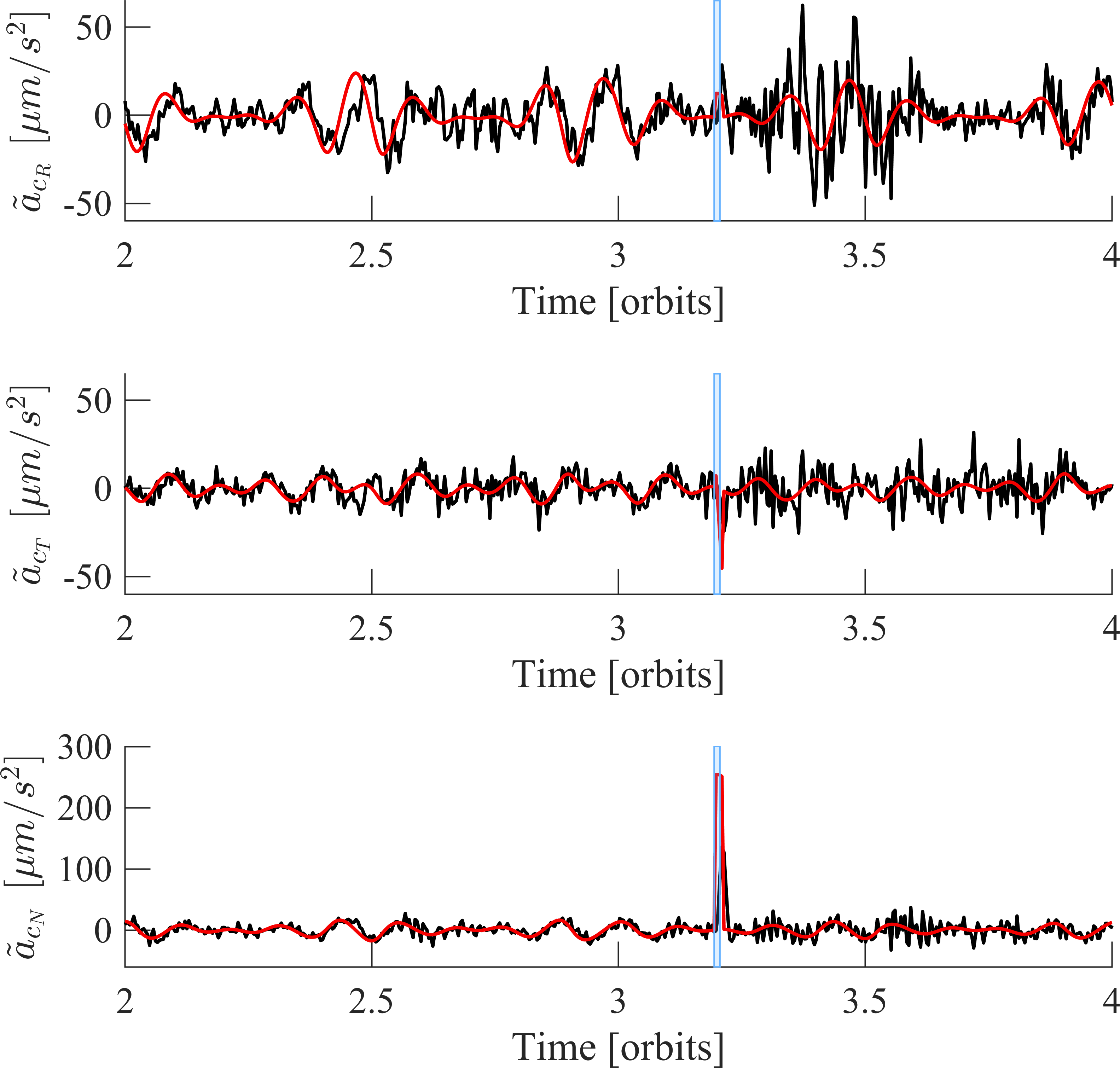}}
\caption{True chief unmodeled accelerations (red) and ADMC estimated empirical accelerations (black) expressed in the chief radial-transverse-normal frame. The radial (top row) and normal (bottom row) axes are aligned with the spacecraft position and angular momentum vectors respectively. The transverse axis (middle row) completes the right-handed triad. The blue region indicates the maneuver duration.}
\label{fig:case study II acceleration tracking}
\end{figure*}

\clearpage
\twocolumn
\section*{Acknowledgment}
This material is based upon work supported by the National Science Foundation Graduate Research Fellowship Program under Grant No. DGE-1656518. Any opinions, findings, and conclusions or recommendations expressed in this material are those of the authors and do not necessarily reflect the views of the National Science Foundation. The authors also wish to thank the Achievement Rewards for College Scientists (ARCS) Foundation for their support. Additionally, this research is part of the Autonomous Nanosatellite Swarming (ANS) Using Radio-Frequency and Optical Navigation project supported by the NASA Small Spacecraft Technology Program cooperative agreement number 80NSSC18M0058.

\ifCLASSOPTIONcaptionsoff
  \newpage
\fi



%
%
%
\bibliographystyle{IEEEtran}
\bibliography{references}   

\begin{IEEEbiography}
[{\includegraphics[width=1in,height=1.25in,clip,keepaspectratio]{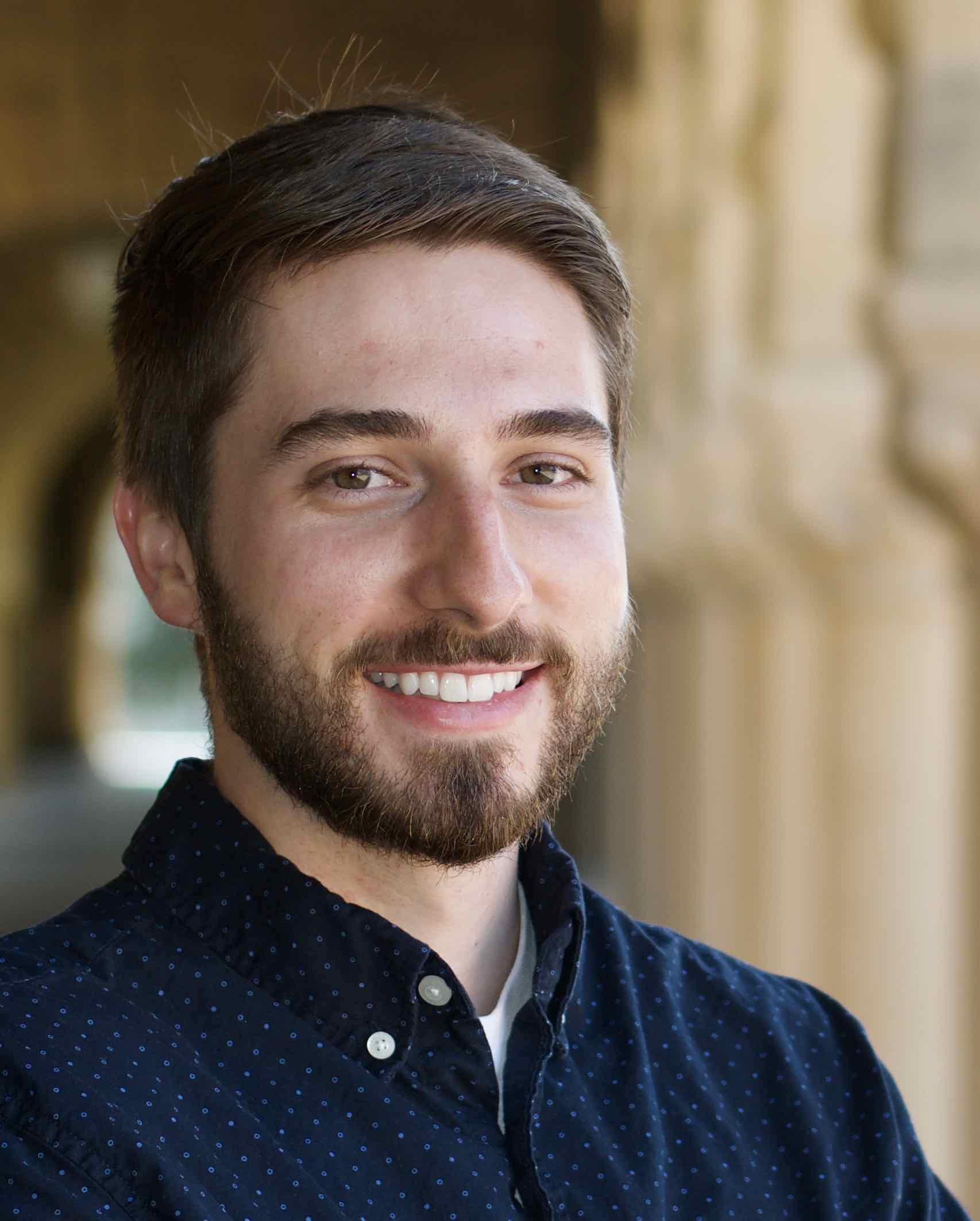}}]
{Nathan Stacey}
received the B.S. degree in mechanical engineering from Utah State University, Logan, UT in 2016. In 2018, he received the M.S. degree in aeronautics and astronautics from Stanford University, Stanford, CA where he is currently pursuing the Ph.D. degree also in aeronautics and astronautics.

Nathan has completed internships with the Northrop Grumman Corporation and Space Dynamics Laboratory. Presently, he is a NASA Pathways intern at Goddard Space Flight Center, Greenbelt, MD and does research in the Stanford Space Rendezvous Laboratory. His research focuses on developing advanced estimation techniques for autonomous orbit determination with application to small celestial body missions. 

Mr. Stacey is the 2016 Utah State University Scholar of the Year, a National Science Foundation Graduate Research Fellow, a Stanford Enhancing Diversity in Graduate Education Doctoral Fellow, and an Achievement Rewards for College Scientists Scholar.
\end{IEEEbiography}

\begin{IEEEbiography}
[{\includegraphics[width=1in,height=1.25in,clip,keepaspectratio]{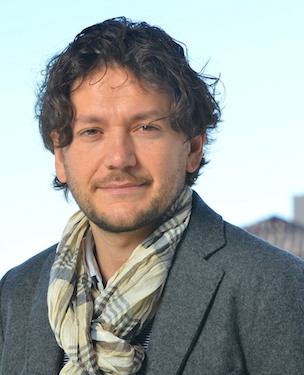}}]
{Simone D'Amico} 
is Associate Professor of Aeronautics and Astronautics at Stanford University. He received the B.S. and M.S. degrees from Politecnico di Milano (2003) and the Ph.D. degree from Delft University of Technology (2010). From 2003 to 2014, he was research scientist and team leader at the German Aerospace Center (DLR). There, he gave key contributions to the design, development, and operations of spacecraft formation-flying and rendezvous missions such as GRACE (United States/Germany), TanDEM-X (Germany), PRISMA (Sweden/Germany/France), and PROBA-3 (ESA). From 2014 to 2020, he was Assistant Professor of Aeronautics and Astronautics at Stanford University. He is the Founding director of the Space Rendezvous Laboratory (SLAB), and Satellite Advisor of the Student Space Initiative (SSSI), Stanford’s largest undergraduate organization. He has over 200 scientific publications and 3000 google scholar’s citations, including conference proceedings, peer-reviewed journal articles, and book chapters. D'Amico's research aims at enabling future miniature distributed space systems for unprecedented science and exploration. His efforts lie at the intersection of advanced astrodynamics, GN\&C, and space system engineering to meet the tight requirements posed by these novel space architectures. The most recent mission concepts developed by Dr. D'Amico are a miniaturized distributed occulter/telescope (mDOT) system for direct imaging of exozodiacal dust and exoplanets and the Autonomous Nanosatellite Swarming (ANS) mission for characterization of small celestial bodies. D’Amico’s research is supported by NASA, NSF, AFRL, AFOSR, KACST, and Industry. He is Chairman of the NASA's Starshade Science and Technology Working Group (TSWG). He is member of the advisory board of space startup companies and VC edge funds. He is member of the Space-Flight Mechanics Technical Committee of the AAS, Associate Fellow of AIAA, Associate Editor of the AIAA Journal of Guidance, Control, and Dynamics and the IEEE Transactions of Aerospace and Electronic Systems. He is Fellow of the NAE’s US FOE Symposium. Dr. D’Amico was recipient of the Leonardo 500 Award by the Leonardo Da Vinci Society and ISSNAF (2019), the Stanford’s Introductory Seminar Excellence Award (2019 and 2020), the FAI/NAA‘s Group Diploma of Honor (2018), the Exemplary System Engineering Doctoral Dissertation Award by the International Honor Society for Systems Engineering OAA (2016), the DLR’s Sabbatical/Forschungssemester in honor of scientific achievements (2012), the DLR’s Wissenschaft Preis in honor of scientific achievements (2006), and the NASA’s Group Achievement Award for the Gravity Recovery and Climate Experiment, GRACE (2004).
\end{IEEEbiography}






\end{document}